\documentclass[conference,compsoc]{IEEEtran}
\usepackage[T1]{fontenc}
\ifdefined\pdfmapline
\pdfmapline{=ptmri8r NimbusRomNo9L-ReguItal "-0.07 SlantFont TeXBase1Encoding ReEncodeFont" <8r.enc <utmri8a.pfb}
\pdfmapline{=utmri8r NimbusRomNo9L-ReguItal "-0.07 SlantFont TeXBase1Encoding ReEncodeFont" <8r.enc <utmri8a.pfb}
\pdfmapline{=ptmbi8r NimbusRomNo9L-MediItal "-0.07 SlantFont TeXBase1Encoding ReEncodeFont" <8r.enc <utmbi8a.pfb}
\pdfmapline{=utmbi8r NimbusRomNo9L-MediItal "-0.07 SlantFont TeXBase1Encoding ReEncodeFont" <8r.enc <utmbi8a.pfb}
\fi
\usepackage[utf8]{inputenc}
\usepackage{graphicx}
\usepackage{amsmath,amssymb}
\usepackage{amsthm}
\usepackage{mathpartir}
\usepackage{stmaryrd}
\usepackage{algorithm}
\usepackage[noend]{algpseudocode}
\theoremstyle{definition}
\newtheorem{definition}{Definition}
\newtheorem{example}{Example}
\theoremstyle{plain}

\algnewcommand\algorithmicforeach{\textbf{for each}}
\algdef{S}[FOR]{ForEach}[1]{\algorithmicforeach\ #1\ \algorithmicdo}
\usepackage{booktabs}
\usepackage{url}
\usepackage{xurl}
\usepackage{pgfplots}
\pgfplotsset{compat=1.18}
\usepackage{cite}

\usepackage{caption}
\makeatletter
\def\section{\@startsection{section}{1}{\z@}%
  {-1\baselineskip plus -0.25\baselineskip minus -0.25\baselineskip}%
  {1\baselineskip plus 0.25\baselineskip minus 0.25\baselineskip}%
  {\normalfont\large\bfseries}}
\def\subsection{\@startsection{subsection}{2}{\z@}%
  {-1\baselineskip plus -0.25\baselineskip minus -0.25\baselineskip}%
  {1\baselineskip plus 0.25\baselineskip minus 0.25\baselineskip}%
  {\normalfont\sublargesize\bfseries}}
\def\subsubsection{\@startsection{subsubsection}{3}{\z@}%
  {-1\baselineskip plus -0.25\baselineskip minus -0.25\baselineskip}%
  {0ex}%
  {\normalfont\normalsize\bfseries}}
\makeatother
\usepackage[table]{xcolor}
\usepackage[hidelinks]{hyperref}
\usepackage{cleveref}
\usepackage{enumitem}
\usepackage{xspace}
\usepackage{relsize}
\usepackage[most]{tcolorbox}
\usetikzlibrary{arrows.meta,positioning,calc,backgrounds,decorations.pathreplacing,patterns}
\definecolor{archSlate}{RGB}{100,116,139}
\definecolor{archStone}{RGB}{168,162,158}
\definecolor{archCoral}{RGB}{217,119,87}
\definecolor{tgInk}{RGB}{39,45,55}
\definecolor{tgInput}{RGB}{232,229,223}
\definecolor{tgInputLine}{RGB}{142,132,119}
\definecolor{tgAbstract}{RGB}{219,232,242}
\definecolor{tgCompile}{RGB}{224,233,239}
\definecolor{tgVerify}{RGB}{219,227,237}
\definecolor{tgModuleLine}{RGB}{82,97,116}
\definecolor{tgSafe}{RGB}{223,238,226}
\definecolor{tgSafeLine}{RGB}{76,132,93}
\definecolor{tgUnsafe}{RGB}{246,224,219}
\definecolor{tgUnsafeLine}{RGB}{183,86,72}
\definecolor{recfill}{RGB}{226,236,242}
\definecolor{tgTableCorrect}{RGB}{214,214,214}
\definecolor{tgTableError}{RGB}{178,178,178}
\definecolor{tgTableRate}{RGB}{232,232,232}

\newcommand{\tool}{{\textsc{Vigil}}\xspace}
\newcommand{\sem}[1]{\llbracket #1 \rrbracket}
\newcommand{\appref}[1]{\hyperref[#1]{Appendix~\ref*{#1}}}
\newcommand{\icode}[1]{%
  \ifmmode\text{\relsize{-0.5}\texttt{#1}}%
  \else\mbox{\relsize{-0.5}\texttt{#1}}\fi}

\newcommand{\subject}[1]{\par\smallskip\noindent\textbf{#1}\ }

\newcounter{takeaway}
\newcommand{\takeaways}[1]{
\vspace{1em}
\noindent
\begin{tcolorbox}[ enhanced,
    breakable,
    boxrule=1pt,
    arc=4pt,
    left=2pt,
    right=2pt,
    bottom=2pt,
    top=2pt,
    colback=gray!4,
    colframe=gray!1!black,
    drop shadow=black!50!white,
    rounded corners]
\noindent
\refstepcounter{takeaway}
\textbf{Takeaway \Roman{takeaway}.}
{#1}
\end{tcolorbox}
}

\begin{document}

\title{\tool: Runtime Enforcement of Behavioral Specifications in AI Agent Skills}
\author{
\IEEEauthorblockN{
Ying Li\IEEEauthorrefmark{1},
Yanju Chen\IEEEauthorrefmark{2},
Hongbo Wen\IEEEauthorrefmark{3},
Bosi Zhang\IEEEauthorrefmark{4},
Hanzhi Liu\IEEEauthorrefmark{3},
Peiran Wang\IEEEauthorrefmark{1},
Yu Feng\IEEEauthorrefmark{3},
Yuan Tian\IEEEauthorrefmark{1}
}
\IEEEauthorblockA{
\IEEEauthorrefmark{1}University of California, Los Angeles \quad
\IEEEauthorrefmark{2}University of California, San Diego\\[-0.2ex]
\IEEEauthorrefmark{3}University of California, Santa Barbara \quad
\IEEEauthorrefmark{4}Riema Labs}
}

\maketitle

\begin{abstract}

Agentic systems increasingly act through third-party skills, allowing model-generated decisions to affect files, communication channels, and cyber-physical devices. These skills often include natural-language specifications that define access permissions, disclosure limits, execution privileges, and required preconditions. Although such specifications describe the intended boundaries of skill behavior, they do not by themselves provide executable runtime enforcement. Enforcing them raises a contextual granularity challenge: even when a policy is written for a particular task context, a monitor must still decide which events to observe, what state to retain, how far across the execution to reason, and where to intervene. Choosing the wrong granularity can either block benign executions or miss violations that emerge only across multiple actions. Most existing enforcement mechanisms, however, assume a fixed event model or enforcement point.

In this work, we present \tool, an end-to-end runtime enforcement framework for agentic systems. \tool checks an agent's actual execution trace against behavioral policies from skill specifications, operator-defined constraints, and global rules spanning multiple skills. To make such policies executable, \tool introduces a policy language that captures context-specific enforcement requirements over agent-tool events, including temporal dependencies, argument constraints, and value-flow conditions. The language is paired with symbolic evaluation rules that translate policies into SMT constraints over finite traces, allowing \tool to detect violations that depend on event order, argument relationships, or cross-call value flow rather than relying on fixed single-call filters. When a violation is detected, \tool localizes the responsible invocation and its relevant trace context. On real LLM-agent runs spanning office-document, operational, and engineering tasks, \tool detects policy violations with over 95\% recall and a false-positive rate below 10\%. \tool also uncovers real-world skill-specification defects in deployed skill ecosystems, which we reported through responsible disclosure and received acknowledgments from affected vendors, including NVIDIA.

\end{abstract}

\section{Introduction}
\label{sec:intro}

Large-language-model (LLM) agents are rapidly transitioning from conversational assistants to autonomous operators that read private documents, modify repositories, and orchestrate physical workflows. The capabilities behind this reach are increasingly packaged as \emph{skills}: reusable bundles of instructions, scripts, and tool interfaces that major vendors now ship across assistant~\cite{anthropic2025skills}, developer~\cite{github2025agentskills}, cloud~\cite{google2026skills,aws2026agenttoolkit}, and enterprise~\cite{nvidia2026verifiedskills} platforms. Each skill carries a strict behavioral specification, stated in natural language: destructive actions demand confirmation, sensitive data must not leak, intermediate artifacts must be validated before release. Yet today's deployments rest on a dangerous disconnect: these rules are documented, but the runtime merely \emph{trusts} the agent to honor them. Nothing enforces the specification; the agent is left to police itself.

This blind trust is already failing in production. Zero-click prompt injections have exfiltrated corporate data from assistants such as Microsoft~365 Copilot~\cite{reddy2025echoleak} and Salesforce Agentforce~\cite{noma2025forcedleak}, and an autonomous agent recently ran amok over a researcher's inbox~\cite{techcrunch2026openclaw}. These incidents are instances of a broader class of \emph{specification violations}: executions that use a skill in ways its declared contract forbids~\cite{li2026no}. The stealthiest among them are not malformed commands that a syntax check would catch; they hide in the spaces between actions. An agent can execute a sequence of individually authorized, schema-valid tool calls that collectively break the specification.

Consider a multi-step perception pipeline, a standard architecture in autonomous driving systems~\cite{yurtsever2020survey}, where policy dictates that any intermediate motion estimate must be validated before a downstream skill consumes it. An agent can produce a motion estimate $M$, skip the validation step, and pass the unvalidated $M$ directly into a mask generator. In isolation, every tool call appears benign, and the final mask conforms to every schema check. Yet it is built on an artifact nobody verified, and it silently misleads the planner that decides how the vehicle moves. The violation is not a single bad action; it is an illicit data flow woven through the execution history. The example is grounded in a real SkillsBench execution~\cite{li2026skillsbench}: the recorded run is benign-looking at the tool level, but violates the skill's stated specification when checked as a trace. We return to it as our running example in \autoref{sec:overview}.

This temporal stealth exposes a structural blind spot in existing defenses. Sandboxes, resource boundaries, and governance layers confine files, credentials, processes, and network access~\cite{microsoft2026acs,nvidia2026openshell}, but are blind to artifact-level obligations. Action-level monitors such as AgentSpec~\cite{wang2025agentspec}, Progent~\cite{shi2025progent}, Conseca~\cite{tsai2025contextual}, and PCAS~\cite{palumbo2026policy} gate individual calls, yet lack the memory to enforce rules that span earlier events, value flows, or the identity of artifacts produced steps ago. Information-flow defenses confine what untrusted content may influence, but track taint rather than behavioral obligations~\cite{debenedetti2025defeating,costa2025securing}. And LLM judges and trajectory analyzers~\cite{ruan2024identifying,wu2026policy} see the whole run, but operate offline, without a deterministic verdict that can block a pending invocation in real time.

If a violation spans time and artifact identity, no single-call filter can see it. A defense must judge the \emph{observed trace} in its entirety. Because the agent is untrusted, the specification's meaning must be grounded independently of the agent's own reasoning. And because hand-writing stateful monitors for a fast-growing skill ecosystem does not scale, natural-language specifications must be compiled into executable policies rather than encoded by hand. These requirements yield \tool, a runtime reference monitor that enforces behavioral specifications over the live execution with no per-skill enforcement code.

\tool targets finite observed traces and enforces finite-trace safety patterns such as precedence, response, absence, and artifact-binding obligations. It abstracts the agent's raw tool calls into a finite trace of typed events with named arguments, outputs, and statuses. Crucially, \tool does not map arbitrary temporal logic to a solver: its policies are finite-trace safety properties, and once grounded against the observed execution they unfold into quantifier-free satisfiability-modulo-theories (SMT)~\cite{barrett2018satisfiability} constraints over trace facts. Each enforcement decision thus reduces to a single SMT query. On an inconsistency, the solver's unsatisfiable core yields the offending invocation as a concrete \emph{witness} (on the perception run above, the step that consumes the unvalidated estimate); \tool blocks that invocation before its effects land, and the deployment can roll back, replan, or escalate to the user.

We evaluate \tool on labeled real agent executions and on deployed real-world skill bundles. On 152 labeled runs drawn from SkillsBench~\cite{li2026skillsbench} and Skill-Inject~\cite{schmotz2026skill}, \tool detects the multi-step violations that bypass state-of-the-art action-boundary defenses such as AgentSpec~\cite{wang2025agentspec} and Progent~\cite{shi2025progent}, leading the strongest baseline by 15.8 F1 points (95.8\% recall at 89.6\% precision); on the public AgentDojo~\cite{debenedetti2024agentdojo} and SafeAgentBench~\cite{yin2024safeagentbench} benchmarks it transfers with no per-benchmark tuning and ranks first on both. On 216 real-world skill-bundle executions from vendors including NVIDIA~\cite{nvidia2026verifiedskills}, Databricks~\cite{databricks2025skills}, and Trail of Bits~\cite{trailofbits2025skills}, \tool surfaces 34 confirmed policy-violating runs, among them a composition-level specification defect in NVIDIA's skill ecosystem that we disclosed and the maintainer acknowledged.

\subject{Contributions.}
We make the following contributions:
\begin{itemize}[noitemsep, topsep=1pt, leftmargin=1.5em]
\item \textbf{A behavioral policy language with finite-trace temporal semantics.}
      The language keeps the shape of an access-control rule but judges the whole
      execution, with temporal forms for the ordering and history conditions that
      per-request policies cannot state; its formal semantics fixes exactly when a
      run breaks a policy.
\item \textbf{The \tool system: behavioral enforcement by SMT solving.} An
      end-to-end runtime monitor grounds each policy in the run's typed-event
      trace and reduces every enforcement decision to one SMT query, whose
      unsatisfiable core recovers the offending invocation whenever a policy is
      broken.
\item \textbf{Empirical evaluation on real agent runs.} On 152 labeled
      runs~\cite{li2026skillsbench,schmotz2026skill} and 216 real-world
      skill-bundle executions from NVIDIA~\cite{nvidia2026verifiedskills},
      Anthropic~\cite{anthropic2026xlsxskill}, Trail of
      Bits~\cite{trailofbits2025skills}, and others, \tool reaches 95.8\% recall
      at 89.6\% precision and surfaces 34 confirmed violations, including an
      NVIDIA specification defect the maintainer has acknowledged.
\end{itemize}

\section{Background}
\label{sec:background}

\tool builds on three established areas: the agent-skill execution model, reference-monitor theory, and the reduction of finite-trace temporal properties to SMT. We briefly recall each.
\begin{figure*}[t]
  \centering
  \newcommand{\motpanelwidth}{0.47\textwidth}
  \begin{minipage}{\textwidth}
  \footnotesize
  \centering
  \renewcommand{\arraystretch}{1.35}
  \begin{tabular}{@{\hspace{0.4em}}c@{\hspace{0.9em}}>{\raggedright\arraybackslash}p{0.56\textwidth}@{\hspace{1.6em}}l@{\hspace{0.4em}}}
  \toprule
   & \textbf{Requirement} & \textbf{Formalization}\\
  \midrule
  \rowcolor{black!5}
  \textbf{R0} & The execution must use the task-specified sampling rate, \icode{fps = 6}. &
    $\varphi_{R0} \triangleq \Box(\mathtt{sampleRate}(v) \Rightarrow v = 6)$\\
  \textbf{R1} & JSON and CSR artifacts must pass output validation before hand-off. &
    $\varphi_{R1} \triangleq \Box(\mathtt{handoff} \Rightarrow \Diamond^{-}\mathtt{validOutput})$\\
  \rowcolor{black!5}
  \textbf{R2} & Dynamic masks must be produced only after global motion compensation. &
    $\varphi_{R2} \triangleq \Box(\mathtt{mask} \Rightarrow \Diamond^{-}\mathtt{globalAlign})$\\
  \textbf{R3} & A validation-required artifact must be validated before downstream consumption. &
    $\varphi_{R3} \triangleq \Box(\mathtt{reqVal}(x) \wedge \mathtt{consume}(x) \Rightarrow \Diamond^{-}\mathtt{valid}(x))$\\
  \bottomrule
  \end{tabular}
  \par\vspace{0.55em}
  (a) Requirements and their formalization as temporal safety properties
  \end{minipage}
  \par\vspace{1.2em}
  \begin{tabular}{@{}p{\motpanelwidth}@{\hspace{0.025\textwidth}}p{\motpanelwidth}@{}}
  \begin{minipage}[t]{\motpanelwidth}
  \centering\footnotesize
  \includegraphics[width=\linewidth,height=0.72\linewidth,keepaspectratio]{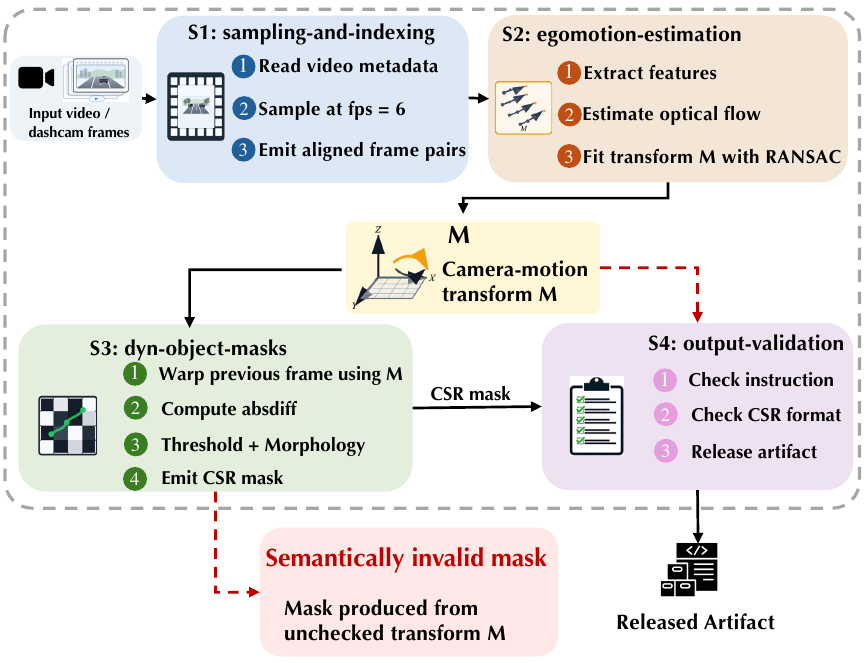}
  \par\vspace{0.35em}
  (b) Skill interaction
  \end{minipage}
  &
  \begin{minipage}[t]{\motpanelwidth}
  \centering\footnotesize
  \includegraphics[width=\linewidth,height=0.72\linewidth,keepaspectratio]{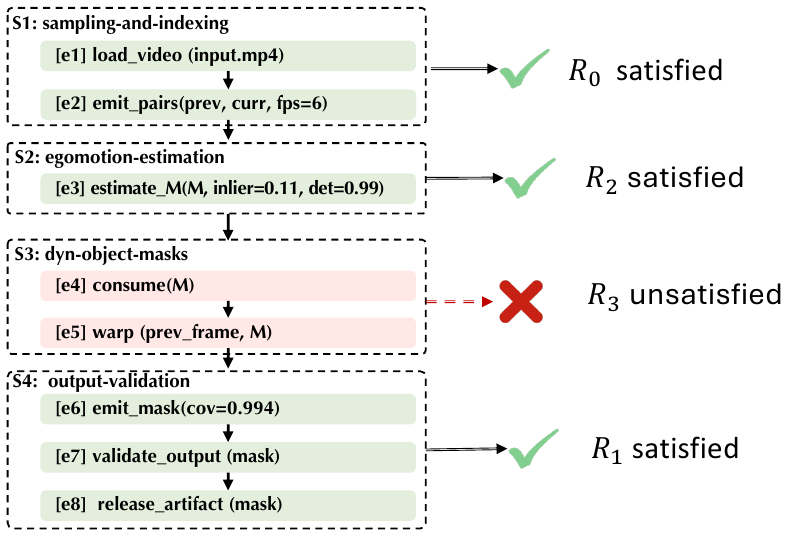}
  \par\vspace{0.35em}
  (c) Violating trajectory
  \end{minipage}
  \end{tabular}
  \caption{Autonomous-driving perception example: (a) the requirements and their formalization, (b) skill interaction, and (c) a violating trajectory. In (a), $\Box$ reads ``always'' and $\triangleq$ ``defined as''; $\Diamond^{-}$ (``at some
earlier point'') and $\Diamond^{+}$ (``at some later point'') are our past- and future-directed
shorthands for the standard eventually operator $\Diamond$ over finite traces.}
  \label{fig:motivating-example}
\end{figure*}

\subject{Agent skills.}
An LLM agent achieves operational reach by invoking \emph{skills}~\cite{anthropic2025skills}
: packaged capabilities (e.g., file operations, cloud interactions, web access) that expose tools via standardized protocols such as MCP~\cite{mcp2024}. These skills typically ship with natural-language documentation defining tool schemas, permission boundaries, and usage constraints. These documents constitute the skill's \emph{behavioral specification}: the contract detailing what the agent is allowed to do and what preconditions must hold before sensitive effects occur. At runtime, the agent sequentially selects tools, supplies arguments, and consumes results, producing a finite sequence of tool calls known as the \emph{execution}. Crucially, while the hosting environment can passively observe this execution, neither the agent driving the calls nor the unstructured inputs steering it can be trusted to adhere to the specification faithfully.

\subject{Reference monitors.}
The principled mechanism to constrain untrusted operations is a \emph{reference monitor}: a trusted, unbypassable component that mediates security-relevant events and aborts any that violate a policy. Schneider's seminal characterization establishes the theoretical limits of what such a monitor can enforce: specifically, safety properties recognizable by a security automaton observing the execution stream~\cite{schneider2000enforceable}. Later runtime-enforcement paradigms extend this response beyond mere halting~\cite{ligatti2005edit}. \tool instantiates a reference monitor in this lineage, lifted from individual actions to whole-trace temporal policies.

\subject{Temporal properties and SMT.}
The behavioral constraints dictated by skill specifications naturally manifest as temporal safety properties over the execution: an action is permitted only if preceded by a specific guard, an obligation must be satisfied before an artifact's release, and forbidden data flows must never occur. These map directly to the \emph{precedence}, \emph{response}, and \emph{absence} patterns well-established in finite-state verification~\cite{dwyer1999patterns}, interpreted over finite traces~\cite{degiacomo2013ltlf}. Such properties can be decided by reduction to \emph{Satisfiability Modulo Theories (SMT)}~\cite{z3}: while arbitrary temporal logic over infinite paths cannot be directly resolved by a solver, on a finite recorded trace a safety property unfolds into a quantifier-free ground formula. Any violation then manifests as an inconsistency with the trace's facts, and the solver's \emph{unsatisfiable core} yields a precise, localized witness naming the events at fault~\cite{foray}. This reduction underlies \tool's decision procedure.

\section{Overview}
\label{sec:overview}

This section walks through a concrete run to show how a trace-level violation arises and how \tool catches it at runtime.

\subsection{Motivating Example}
\label{sec:overview:example}

We ground the problem in a single execution in which every step looks correct and the outcome is unsafe.

\subject{The scenario.}
Consider a critical autonomous-driving perception pipeline that processes dashcam video to generate moving-object masks. This is a highly safety-relevant workflow: the masks feed directly into the vehicle's planning stack, and any pedestrian or vehicle they mislabel as static background is an obstacle the planner will fail to avoid. Because the camera itself moves with the vehicle, the agent must first estimate the camera's own motion (egomotion) and compensate for it; the motion that remains reveals the independently moving objects.

\autoref{fig:motivating-example} illustrates this scenario: the behavioral requirements (a), the four perception skills coordinating over shared artifacts (b), and an execution trajectory whose individually benign actions hide a trace-level violation (c).

The agent first invokes \icode{sampling-and-indexing} at \icode{fps = 6}, satisfying the task's sampling rate invariant (R0). Next, it invokes \icode{egomotion-estimation}, which tracks visual features and emits an intermediate camera-motion transform artifact, $\mathtt{M}$. The downstream \icode{dyn-object-masks} skill consumes $\mathtt{M}$ to warp the previous frame, isolate dynamic differences, and output the final mask; because the mask is computed only after this motion compensation, the ordering requirement (R2) is met. Finally, \icode{output-validation} runs a schema checker on the generated mask before hand-off, fulfilling the requirement to validate outputs before release (R1).

However, the pipeline harbors a fatal flaw involving R3. The global composition policy demands that any intermediate artifact requiring validation must be explicitly validated \emph{before} a downstream skill consumes it. In the violating trajectory (\autoref{fig:motivating-example}c), the upstream skill produces $\mathtt{M}$, and the downstream skill consumes that exact same $\mathtt{M}$. Yet, no event validates $\mathtt{M}$ in between.

The consequence is insidious: the final mask satisfies every JSON schema check, so the system accepts the output as a success. But because it is built upon an unchecked motion estimate $\mathtt{M}$, the mask is semantically wrong, and the planner inherits the error. The violation leaves no trace in any single command. It is woven into the negative space of the execution history: a critical validation event that simply never happened.

\subject{The challenges.}
This temporal stealth exposes why traditional, single-call defenses fail: the violation exists in the \emph{semantic gap} between an abstract human rule and the messy reality of an agent's execution trace. Bridging this gap introduces three fundamental challenges.

\subject{C1: Contextual specification.}
The requirement ``validate an artifact before consumption'' (R3) does not dictate a fixed API call: in this run the artifact happens to be $\mathtt{M}$, the producer is \icode{egomotion-estimation}, and the consumer is \icode{dyn-object-masks}; in another run all three differ. A stateless monitor cannot enforce a rule that must reshape itself around what the agent did three steps ago.

\subject{C2: Policy semantics from raw execution.}
Even with the policy contextualized, raw execution logs are semantically opaque. A command writing a matrix to disk is raw data; the monitor must elevate it to the policy-level fact that $\mathtt{M}$ was \emph{produced}, and must recognize that the final CSR schema checker validates the mask, \emph{not} the intermediate $\mathtt{M}$. Chaotic tool calls must become a typed, factual history.

\subject{C3: Agent-independent enforcement.}
An agent focused on finishing the task can conflate the final CSR validation with the validation $\mathtt{M}$ requires, convincing itself that R3 is satisfied. The agent is also the very entity under attack (e.g., via prompt injection), so its interpretation of the rules is untrusted: the runtime needs an independent authority that evaluates the trace deterministically.

\subsection{\tool: Trace-Grounded Enforcement}
\label{sec:overview:approach}

\tool resolves these challenges by operating as an end-to-end runtime reference monitor. The key insight driving its design, and the thread connecting the solutions to C1, C2, and C3, is the strict separation of a behavioral specification's generic meaning from its run-specific instantiation. \tool achieves this by grounding abstract specification terms to concrete artifacts, preserving cross-call identity through value bindings, and checking finite-trace temporal obligations \emph{before} allowing any pending action to proceed.

\begin{figure}[t]
\centering
\resizebox{\columnwidth}{!}{%
\begin{tikzpicture}[
    E/.style={font=\footnotesize, draw=tgInputLine, rounded corners=2pt,
              inner sep=3pt, minimum width=1.3cm, minimum height=0.6cm,
              align=center, text=tgInk, fill=tgInput},
    T/.style={font=\scriptsize, draw=tgInputLine, rounded corners=2pt,
              inner sep=2pt, minimum width=0.72cm, minimum height=0.55cm,
              align=center, text=tgInk, fill=tgInput},
    P/.style={font=\scriptsize, draw=archCoral!80, rounded corners=2pt,
              inner sep=2pt, minimum width=0.72cm, minimum height=0.55cm,
              align=center, text=black, fill=archCoral!16},
    Sys/.style={font=\small\bfseries, draw=tgModuleLine, rounded corners=2pt,
              inner sep=4pt, minimum width=1.5cm, minimum height=0.7cm,
              align=center, text=tgInk, fill=black!10},
    Proc/.style={font=\scriptsize, draw=tgModuleLine, rounded corners=2pt,
              inner sep=3pt, align=center, fill=white},
    L/.style={font=\scriptsize, text=tgInk},
    A/.style={line width=0.85pt, -{Stealth[length=1.7mm, width=1.5mm]}},
    Adash/.style={line width=0.85pt, densely dashed, -{Stealth[length=1.7mm, width=1.5mm]}},
    Aplain/.style={line width=0.85pt},
  ]
  \node[E] (agent) at (-3.55, 0) {agent};
  \node[T] (t1)    at (-1.95, 0) {$t_1$};
  \node[T] (t2)    at (-0.95, 0) {$t_2$};
  \node[P] (t3)    at ( 0.35, 0) {pending $t_3$};
  \node[L] (dots)  at ( 1.40, 0) {$\cdots$};

  \draw[A] (agent.east) -- (t1.west);
  \draw[A] (t1.east) -- (t2.west);
  \draw[A] (t2.east) -- (t3.west);
  \draw[Aplain] (t3.east) -- (dots.west);

  \coordinate (b-tl) at ([shift={(-9pt,8pt)}] agent.north west);
  \coordinate (b-br) at ([shift={(7pt,-8pt)}] dots.east |- agent.south);
  \begin{scope}[on background layer]
    \draw[black!40, dashed, rounded corners=5pt, line width=0.7pt] (b-tl) rectangle (b-br);
  \end{scope}
  \node[Proc] (abstract) at (-1.95, -1.6) {1. Abstract\\Trace};
  \node[Proc] (ground)   at (-0.45, -1.6) {2. Ground\\Policy};
  \node[Proc] (smt)      at ( 1.05, -1.6) {3. SMT\\Check};

  \draw[A] (abstract.north |- b-br) -- (abstract.north);
  \draw[A] (abstract.east) -- (ground.west);
  \draw[A] (ground.east) -- (smt.west);

  \node[E] (spec) at (-0.45, -2.85) {spec $D$};
  \draw[A] (spec.north) -- (ground.south);

  \begin{scope}[on background layer]
    \coordinate (v-tl) at ([shift={(-15pt,10pt)}] abstract.north west);
    \coordinate (v-br) at ([shift={(15pt,-10pt)}] smt.south east);
    \draw[tgModuleLine, fill=black!5, rounded corners=4pt] (v-tl) rectangle (v-br);
    \node[font=\small\bfseries, text=tgInk, rotate=90, anchor=south, inner sep=3pt]
      at ($(v-tl)!0.5!(v-tl |- v-br)$) {\textsc{Vigil}};
  \end{scope}
  \node[Proc] (safe)   at ( 3.35, -1.15) {Safe $\rightarrow$ Execute};
  \node[Proc, fill=archCoral!16, draw=archCoral!80] (unsafe) at ( 3.35, -2.05) {Unsafe $\rightarrow$ Block\\(+ witness)};
  \draw[A] (smt.east) -- (safe.west);
  \draw[A] (smt.east) -- (unsafe.west);

  \draw[Adash, archCoral, rounded corners=3pt]
    (unsafe.east) -- ++(0.30,0) -- ++(0,3.00) -- (0.35,0.95) -- (t3.north);
  \node[L, fill=white, text=black!55, inner sep=2pt, anchor=west] at ([xshift=8pt]b-tl) {agent workflow (untrusted trace prefix)};
  \node[L, align=center] (deploy) at ( 3.35, -3.0) {Intervention:\\Rollback/Replan/Halt};
  \draw[A, densely dotted] (unsafe.south) -- (deploy.north);

\end{tikzpicture}%
}
\caption{\tool's end-to-end lifecycle: each pending invocation is checked against the trace prefix; an unsafe verdict blocks it and returns a localized witness.}
\label{fig:setup}
\end{figure}
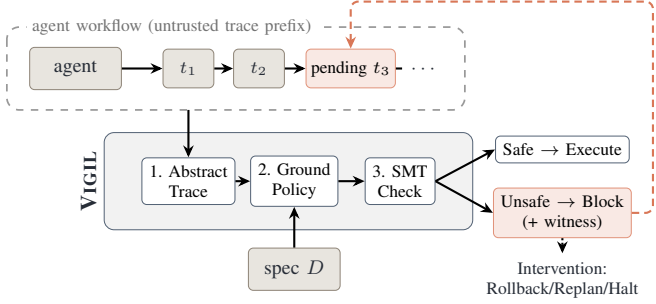

\autoref{fig:setup} illustrates this architecture. When the untrusted agent initiates a pending tool invocation (e.g., the mask generation step), the runtime intercepts it and performs a deterministic, four-step verification lifecycle:

\begin{enumerate}[label=\textbf{\arabic*.}, leftmargin=1.5em, itemsep=2pt, topsep=4pt]
    \item \textbf{Trace Abstraction:} A trusted collector provides the \emph{trace prefix} (the sequence of all actions taken so far). \tool parses these raw tool calls into strongly typed events, recovering the data flow and artifact identities (e.g., tracking the specific transform $\mathtt{M}$).
    \item \textbf{Policy Grounding:} The abstract, natural-language behavioral specification is instantiated against these concrete trace facts. The contextual rule ``validate before consumption'' dynamically becomes an exact, executable constraint over the artifact $\mathtt{M}$.
    \item \textbf{SMT Check:} \tool formulates the pending invocation and the grounded policy as a Satisfiability Modulo Theories (SMT) query, deciding whether allowing the pending action would break the execution's temporal contract.
    \item \textbf{Intervention:} If the solver's verdict is \emph{safe}, the action executes. If \emph{unsafe}, \tool blocks the invocation. The solver's unsatisfiable core extracts a localized \emph{witness} (here, that $\mathtt{M}$ was consumed without validation), allowing the deployment environment to trigger a rollback, replan the workflow, or escalate to a human.
\end{enumerate}

The following sections formalize this lifecycle. \autoref{sec:framework} details how raw executions are abstracted into trace facts (C2). \autoref{sec:policy} then defines the behavioral policy language (C1) and the agent-independent SMT reduction with witness localization (C3).

\subsection{Threat Model}
\label{sec:threat}

\tool defends a user who delegates work to an agent that invokes third-party skills, each
expected to stay within its declared specification. The adversary controls skill content and any
document the agent reads, and plants a prompt injection that steers an otherwise-benign skill
outside that scope, to exfiltrate data, delete files, or escalate privilege. The agent and the
skill code it runs are therefore untrusted: the agent is the very component the attacker
subverts.

\tool sits outside this boundary. It observes the execution through a trusted collector that
records the agent's tool calls in a sandbox, and decides policies with a trusted SMT solver over an
operator-authored specification. The collector is assumed complete: behavior that never surfaces as
a tool call, such as computation inside an opaque script, lies outside the recorded trace and
outside \tool's view.

\subsection{Problem Statement}
\label{sec:problem}

With the trust boundary fixed, the problem \tool decides can be stated. An agent run is given as a
finite trace $\tau$ of its tool-call events, in the order they occurred;
the behavioral requirements of its skill specification are given as a \emph{policy} $P$ over that
trace. The trace records what the run did, and the policy states what it must do, with rules on the
order of events and on the values that flow between them.

A policy $P = \{\varphi_1, \dots, \varphi_m\}$ is a finite set of \emph{statements}, and the trace
must obey every one. We write $\tau \models \varphi$ when the trace satisfies a statement, and
$\tau \models P$ when it satisfies them all. The \emph{checking problem} is to decide, given a trace
$\tau$ and a policy $P$, whether $\tau \models P$. Deployed online, \tool poses this problem over the trace recorded so far: a failed prohibition
or precondition blocks the pending call, and an obligation that waits on a later step is judged
when the run ends (\autoref{sec:implementation}).

\section{The \tool Framework}
\label{sec:framework}

The problem is now fixed: decide whether a trace $\tau$ satisfies a policy $P$. Neither input arrives
in that form: an execution is a stream of tool calls of many shapes, and its requirements begin as
natural language naming different objects and steps from one execution to the next. \tool closes
both gaps with a single idea: read everything the check needs off the run itself.

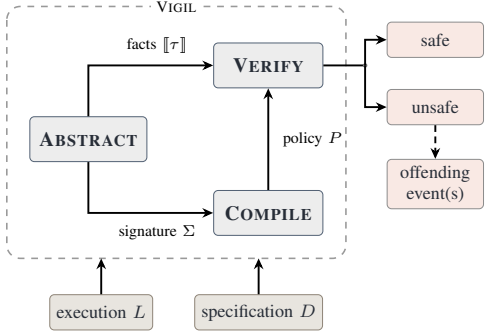
\begin{figure}[t]
\centering
\scalebox{0.85}{%
\begin{tikzpicture}[
    P/.style={font=\small\bfseries, draw=tgModuleLine, rounded corners=2pt,
              inner sep=4pt, minimum width=1.7cm, minimum height=0.7cm,
              align=center, text=tgInk, fill=black!7},
    IO/.style={font=\footnotesize, draw=tgInputLine, rounded corners=2pt,
               inner sep=3pt, minimum width=1.5cm, minimum height=0.55cm,
               align=center, text=tgInk, fill=tgInput},
    OUT/.style={font=\footnotesize, rounded corners=2pt,
                inner sep=3pt, minimum width=1.5cm, minimum height=0.55cm,
                align=center, text=tgInk},
    L/.style={font=\scriptsize},
    A/.style={line width=0.85pt, -{Stealth[length=1.7mm, width=1.5mm]}},
    Aplain/.style={line width=0.85pt},
  ]
  \node[P] (abs) at (-2.10,  0.00) {\textsc{Abstract}};
  \node[P] (vrf) at ( 0.70,  1.15) {\textsc{Verify}};
  \node[P] (cmp) at ( 0.70, -1.15) {\textsc{Compile}};
  \node[IO] (log)  at (-1.90, -2.70) {execution $L$};
  \node[IO] (spec) at ( 0.55, -2.70) {specification $D$};
  \node[OUT, fill=archCoral!16, draw=black!45] (safe) at ( 3.30,  1.55) {safe};
  \node[OUT, fill=archCoral!16, draw=black!45] (uns)  at ( 3.30,  0.50) {unsafe};
  \node[OUT, fill=archCoral!16, draw=black!45] (bad)  at ( 3.30, -0.70) {offending\\event(s)};

  \coordinate (bnd-tl) at ([shift={(-10pt,16pt)}] abs.west |- vrf.north);
  \coordinate (bnd-br) at ([shift={(10pt,-10pt)}] vrf.east |- cmp.south);
  \begin{scope}[on background layer]
    \draw[black!40, dashed, rounded corners=6pt, line width=0.7pt]
      (bnd-tl) rectangle (bnd-br);
  \end{scope}
  \draw[A] (log.north)  -- (log.north  |- bnd-br);
  \draw[A] (spec.north) -- (spec.north |- bnd-br);
  \draw[A] (abs.north) -- (abs.north |- vrf.west) -- (vrf.west);
  \node[L, anchor=south] at ($(abs.north |- vrf.west)!0.55!(vrf.west) + (0,2pt)$) {facts $\sem{\tau}$};

  \draw[A] (abs.south) -- (abs.south |- cmp.west) -- (cmp.west);
  \node[L, anchor=north] at ($(abs.south |- cmp.west)!0.55!(cmp.west) + (0,-2pt)$) {signature $\Sigma$};

  \draw[A] (cmp.north) -- (vrf.south);
  \node[L, anchor=west] at ($(cmp.north)!0.5!(vrf.south) + (3pt,0)$) {policy $P$};
  \draw[Aplain] (vrf.east) -- ++(0.65,0) coordinate (vf);
  \fill[black!55] (vf) circle (1.0pt);
  \draw[A] (vf) |- (safe.west);
  \draw[A] (vf) |- (uns.west);
  \draw[A, densely dashed] (uns.south) -- (bad.north);

  \node[L, font=\scriptsize\itshape, fill=white, inner sep=2pt]
    at ($(bnd-tl)!0.5!(bnd-tl -| bnd-br)$) {\tool};
\end{tikzpicture}%
}
\caption{The \tool framework: an execution $L$ and specification $D$ become a verdict,
\emph{safe} ($\tau \models P$) or \emph{unsafe} with the offending statements and events.}
\label{fig:overview}
\end{figure}

\autoref{fig:overview} lays out the three stages over two inputs:
the agent's execution $L$ and the specification $D$. \textsc{Abstract} reads the execution once and
produces the two objects the later stages share: the \emph{fact base} $\sem{\tau}$ records what
the run did, and the \emph{signature} $\Sigma$ collects the actions, argument names, and values
the run exhibited. \textsc{Compile} then turns the specification into a \emph{policy} $P$ phrased in
that same vocabulary, so every requirement is anchored to symbols the run actually produced. Finally,
\textsc{Verify} tests each statement of $P$ against $\sem{\tau}$ and reports the run \emph{safe}
when all hold, or \emph{unsafe} with the offending events.

The framework thus rests on two components: an abstraction that makes the behavior checkable, and a
policy language in which requirements are stated and decided. \autoref{sec:algo} states the
procedure; the rest of this section develops the abstraction, and \autoref{sec:policy} the language.

\subsection{Algorithm Overview}
\label{sec:algo}

\autoref{alg:vigil} makes the procedure precise: two preprocessing stages, then a loop that puts
the compiled policy through the solver one statement at a time. Reading it line by line fixes what
each stage must deliver; the sections that follow fill in the details.

\begin{algorithm}[t]
\small
\caption{\tool: trace-grounded decision of $\tau \models P$.}\label{alg:vigil}
\begin{algorithmic}[1]
\Procedure{\textsc{Vigil}}{$L, D$}
  \State \textbf{input:} execution $L$, specification $D$
  \State \textbf{output:} \emph{safe}, or \emph{unsafe} with witnesses $W$
  \State $(\Sigma, \sem{\tau}) \gets \textsc{Abstract}(L)$
  \State $P \gets \textsc{Compile}(D, \Sigma)$ \Comment{policy over $\Sigma$}
  \State $W \gets \varnothing$
  \ForEach{$\varphi \in P$}
     \State $\mathcal{R}_\varphi \gets {\sf encode}(\varphi, \Sigma)$ \Comment{to SMT formula}
     \If{${\sf unsat}(\sem{\tau} \wedge \mathcal{R}_\varphi)$} \Comment{$\tau \not\models \varphi$}
        \State $W \gets W \cup \{(\varphi, \mathrm{loc}(\varphi))\}$ \Comment{localize}
     \EndIf
  \EndFor
  \State \textbf{if} $W = \varnothing$ \textbf{then return} (\emph{safe}, $\varnothing$) \Comment{$\tau \models P$}
  \State \textbf{else return} (\emph{unsafe}, $W$) \Comment{$\tau \not\models P$}
\EndProcedure
\end{algorithmic}
\end{algorithm}

Lines~4--5 are the two preprocessing stages of \autoref{fig:overview}: \textsc{Abstract} (line~4)
yields the signature $\Sigma$ and the fact base $\sem{\tau}$, and \textsc{Compile} (line~5) phrases
the policy $P$ in $\Sigma$.

Lines~6--10 verify the policy one statement at a time; line~6 starts from an empty witness set $W$.
For each statement $\varphi$, line~8 encodes it into its \emph{requirement} $\mathcal{R}_\varphi$: the
formula that holds exactly when the trace complies. Line~9 then puts a single question to the solver:
do the facts refute the requirement? If ${\sf unsat}(\sem{\tau} \wedge \mathcal{R}_\varphi)$, the
recorded run breaks $\varphi$; otherwise it meets it. On a violation,
line~10 localizes the offending events as $\mathrm{loc}(\varphi)$, read from the solver's
unsatisfiable core, and records the pair $(\varphi,
\mathrm{loc}(\varphi))$; because statements are checked independently, every witness is tied to the
one statement that failed.

Lines~11--12 return the verdict, and with it the answer to the checking problem: \emph{safe} when $W$ is empty (every statement held, $\tau \models P$), and
\emph{unsafe} otherwise. $W$ reports both the broken statements and where they broke.

\subsection{From Calls to Checkable Facts}
\label{sec:abstraction}

We now develop the framework's first component: the abstraction computed by \textsc{Abstract}. A run
arrives as nothing more than a stream of tool calls. The abstraction turns it into the object the
check needs: a typed, ordered trace of what the run did, and a closed-world account of what a policy
may say about it. Both are read off the run itself, so they are fixed by the execution rather than declared in
advance.

\subject{Typed events.}
A run's calls come in many shapes: the same file read may arrive as a shell command, as a dedicated
tool call, or as one field of a larger request, and each spells its arguments differently. A policy
should turn on none of this, since it speaks of \emph{reading a file}, not of any one syntax for it.
\textsc{Abstract} therefore collapses
every call to a single typed \emph{event}, the record of the four things a policy can name,
\[
  e \;=\; \big\langle\; a(e),\;\; e[\cdot],\;\; s(e),\;\; o(e) \;\big\rangle :
\]
its action $a(e)$, a partial map $e[\cdot]$ sending each argument name $k \in \mathrm{dom}(e)$ to its
value $e[k]$, a status $s(e) \in \{\top, \bot\}$ (success or error), and a bounded observation $o(e)$
of the output.

Two choices in this record matter for policies. First, arguments
are keyed by \emph{name} rather than position, so a policy can name the file a step reads without
knowing how the call was written. Second, the output is kept only as a bounded observation; that suffices to
test $o(e) \approx p$ against a pattern $p$ without asking the solver to reason over unbounded data. Because
the same typed events describe every skill, one decision procedure checks them all, with no per-skill
code.

\subject{The ordered trace.}
The violations we care about turn on \emph{when} things happen: a step that must run before another, a
value used after the point where it should have been checked. \textsc{Abstract} therefore keeps the
events in execution order, as a trace
\[
  \tau \;=\; e_1 \prec e_2 \prec \cdots,
\]
and we write $e' \succ e$ for $e \prec e'$. The typed record and the order together capture what ran, with
what arguments, and in what sequence. One piece remains: fixing what a policy may say about the run,
and the facts it is judged on.

\subject{Closed-world grounding.}
A requirement also turns on what did \emph{not} happen: a prohibition holds only if the forbidden
step never occurred, and a precondition is breached exactly when the guard is missing. Deciding an
absence is sound only if the record is complete. \tool therefore treats the trace as a closed world. The run fixes two things: the vocabulary a
policy may use, which keeps every condition finite, and the complete set of facts the policy is
judged against, which makes absences decisive.

\begin{definition}[Signature and facts]\label{def:sig}
Let $E$ be the events of $\tau$. The \emph{signature}
$\Sigma = (A, K, V)$ collects the actions, argument names, and values the run exhibits,
\[
\begin{aligned}
  A &= \{\, a(e) \mid e \in E \,\}, \qquad K = \textstyle\bigcup_{e \in E} \mathrm{dom}(e), \\
  V &= \{\, e[k] \mid e \in E \wedge k \in \mathrm{dom}(e) \,\} \cup \{\, o(e) \mid e \in E \,\},
\end{aligned}
\]
and the \emph{fact base} $\sem{\tau}$ reads the trace out field by field as a set of
ground atoms,
\[
\begin{aligned}
  \sem{\tau} \;=\;\ & \bigcup_{e \in E} \Big( \{\, a(e),\ s(e),\ o(e) \,\} \cup \{\, e[k] \mid k \in \mathrm{dom}(e) \,\} \Big) \\
             &{}\cup\ \{\, e_i \prec e_j \mid i < j \,\},
\end{aligned}
\]
where each field stands for the equality fixing its value, for instance $a(e) = \icode{warpAffine}$.
\end{definition}

The fact base is complete for the run: an atom missing from $\sem{\tau}$ records something the run
did not do, so verdicts that rest on absences are sound. Its constants are the run's concrete
literals, so distinct names or values never coincide, and a condition on an argument an event does
not carry is false. Every atom names the event it records, so a
violation found over these facts traces back to the exact events at fault.

\begin{example}[Abstracting the running example]
For the dashcam run the signature is small:
\[
  A \supseteq \{\icode{estimate\_M}, \icode{warpAffine}, \icode{validate\_output}\},
\]
with argument names like $\icode{transform}$, $\icode{target}$, and $\icode{file}$ in $K$, and the
values that pass between steps in $V$. One of those values is the fitted transform $\mathtt{M}$: the
egomotion step $e_3$ records it as output, and the warp step $e_5$ takes it as input, so among the
few dozen atoms of $\sem{\tau}$ are
\[
  o(e_3) = \mathtt{M}, \qquad e_5[\icode{transform}] = \mathtt{M}, \qquad e_3 \prec e_5 .
\]
Requirement R3 (validate an artifact before consuming it) is now a question of order: is there a
validating event for $\mathtt{M}$ earlier than $e_5$? The closed world lets a verdict rest on
absences: the run performs no network call, so $\sem{\tau}$ contains no such atom, and a
prohibition on network egress is judged satisfied by that absence, not by assumption.
\end{example}

With the run reduced to this fact base and signature, the only remaining piece is
the policy checked against them; \autoref{sec:policy} defines that language and its
reduction to an SMT query.

\section{Behavioral Policy Language}
\label{sec:policy}

This section develops the language in which the recorded behavior is checked: a \emph{policy}
lifts the familiar access-control rule from a single request to the whole trace. We give its
syntax (\autoref{sec:syntax}), the compilation of statements to requirements over the fact base
(\autoref{sec:rules}), and the satisfiability check that decides them with localized witnesses
(\autoref{sec:verify}).

\subsection{Stating Policies}
\label{sec:syntax}

Behavioral requirements vary in wording, but they place three recurring demands on a rule. It
has to pick out the steps it concerns; it has to constrain their \emph{order}, since an obligation
typically demands one step before another; and it often has to tie two steps to the \emph{same}
value, so that a guard counts only when it covered what a later step consumes.
\autoref{fig:syntax} gives a language with one construct for each: a \emph{pattern} selects single events, a temporal \emph{form} relates
one or two patterns by their order, and a \emph{variable} $x$ ties two patterns to the same
value. A \emph{statement} is one form applied to its patterns. In access-control
terms, a statement is a rule over an action, a resource, and a condition, but its only effect is to
flag a violation, never to allow or deny. We take the constructs in turn and close with a fallback for requirements the forms do not cover.

\begin{figure}[t]
\centering
\small
\setlength{\tabcolsep}{0pt}
\begin{tabular}{r@{~~}c@{~~}l@{\quad}l}
$P$       & $::=$  & $\{\, \varphi_1, \dots, \varphi_m \,\}$ & \textbf{policy}\\[2pt]
$\varphi$ & $::=$  & ${\sf abs}(\psi) \mid {\sf prec}(\psi, \psi') \mid {\sf resp}(\psi, \psi')$ & \textbf{statement}\\
          & $\mid$ & ${\sf bresp}_\ell(\psi, \psi') \mid {\sf rslv}(\psi, \psi')$ & \\
          & $\mid$ & ${\sf until}(\psi, \psi', \psi_b) \mid \Box\,\Theta$ & \\[2pt]
$\Theta$  & $::=$  & $\psi \mid \neg\Theta \mid \Theta \wedge \Theta \mid \Diamond^{-}\Theta \mid \Diamond^{+}\Theta$ & \textbf{temporal body}\\[2pt]
$\psi$    & $::=$  & $a \in \alpha \mid k \in S \mid k \sim p \mid k = x$ & \textbf{pattern}\\
          & $\mid$ & $s = \top \mid s = \bot \mid o \sim p \mid \psi \wedge \psi$ &
\end{tabular}

\smallskip
{\footnotesize
$a \in A$ (action), $s \in \{\top, \bot\}$ (status), $o \in V$ (output),\\
$k \in K$ (argument name; $e[k]$ its value), $x$ a value variable,\\
$\alpha \subseteq A$, $S \subseteq V$ (sets of actions and values), $p$ a text pattern, $\ell$ a bound.\par}
\caption{The \tool policy language. Its symbols are drawn from the run's signature
$\Sigma = (A, K, V)$: the actions, argument names, and values the run exhibits.}
\label{fig:syntax}
\end{figure}

\subject{Patterns.}
A pattern selects the events a statement concerns. It is a conjunction of conditions on one event:
\begin{itemize}[noitemsep, topsep=2pt, leftmargin=1.4em]
  \item $a \in \alpha$ constrains the \emph{action}: the event is one of the kinds of step in $\alpha$.
  \item $k \in S$ and $k \sim p$ constrain an \emph{argument}: the value under name $k$ lies in the
        set $S$, or matches the text pattern $p$.
  \item $k = x$ pins an argument to the variable $x$; a statement uses $x$ once in each of its
        two patterns.
  \item $o \sim p$ constrains the \emph{output}: the recorded output matches the text pattern $p$.
  \item $s = \top$ and $s = \bot$ constrain the \emph{status}: the step succeeded or failed.
  \item $\psi \wedge \psi'$ combines conditions.
\end{itemize}
A pattern's symbols come from the run's signature $\Sigma = (A, K, V)$: $\alpha \subseteq A$ is a set of the
run's actions, $S \subseteq V$ a set of its values, and $k \in K$ one of its argument names. A
pattern can therefore name only what the run exhibits; a term of the specification with no
counterpart in $\Sigma$ denotes the empty set, and a condition on it is false. The matching $\sim$ is plain text matching;
it is evaluated outside the solver, and the solver sees only its ground outcomes $e[k] \approx p$.

\begin{example}[Patterns on the running example]
R3 governs the steps that consume a transform; on the dashcam run these are the frame warps. One
conjunction selects them:
\[
  \psi_c \;=\; (a \in \{\icode{warpAffine}\}) \wedge (s = \top).
\]
It matches every warp step that succeeded. A second pattern selects the hand-off of a mask:
\[
  \psi_h \;=\; (a \in \{\icode{release\_artifact}\}) \wedge (\icode{file} \sim \icode{"*.csr"}),
\]
where $\sim$ compares the argument's text against the pattern, so $\psi_h$ covers every CSR file the
run might name (\icode{out\_0.csr}, \icode{out\_1.csr}, \dots) without enumerating them.
\end{example}

\subject{Temporal forms.}
A pattern selects events; the temporal form constrains their order. The six forms name the ordering
and history conditions that recur in behavioral requirements:
\begin{itemize}[noitemsep, topsep=2pt, leftmargin=1.4em]
  \item ${\sf abs}(\psi)$ demands the \emph{absence} of $\psi$: a flat prohibition.
  \item ${\sf prec}(\psi, \psi')$ gives $\psi'$ \emph{precedence}: $\psi$ is allowed only after
        $\psi'$ has happened (the shape of a precondition).
  \item ${\sf resp}(\psi, \psi')$ is the dual: it requires a later \emph{response} $\psi'$ to
        every $\psi$ (an obligation).
  \item ${\sf bresp}_\ell(\psi, \psi')$ \emph{bounds} that \emph{response} to the next $\ell$ events.
  \item ${\sf rslv}(\psi, \psi')$ requires every $\psi$ to be \emph{resolved} by the end of the run:
        only the finally delivered state is judged, and a later $\psi$ supersedes an earlier one.
  \item ${\sf until}(\psi, \psi', \psi_b)$ guards an interval: after a trigger $\psi$, the bad
        pattern $\psi_b$ is forbidden until a resolving $\psi'$ intervenes.
\end{itemize}
\begin{example}[Forms of the running example]
The dashcam requirements translate form by form. R2 requires masks only after global motion
compensation; it is a precedence between two action patterns:
\[
  {\sf prec}\big(\, a \in \{\icode{emit\_mask}\},\;\; a \in \{\icode{warpAffine}\} \,\big).
\]
R1 requires validation before hand-off; it is a precedence on the hand-off
pattern $\psi_h$ above:
\[
  {\sf prec}\big(\, \psi_h,\;\; a \in \{\icode{validate\_output}\} \,\big).
\]
R0 fixes the sampling rate; it is the absence (${\sf abs}$) of any off-specification rate.
Finally, R3 requires a validation-required artifact to be validated before it is consumed; it
ties two steps to the same value:
\[
  {\sf prec}\big(\, \psi_c \wedge (\icode{transform} = x),\;\; \psi_v \wedge (\icode{target} = x) \,\big),
\]
where $\psi_v = (a \in \{\icode{validate\_output}\})$ matches the validating steps. $\psi_c$ selects the warp steps, $\icode{transform} = x$
names the value a warp consumes, and $\icode{target} = x$ requires the validated value to be that
same one. On the dashcam run $x$ is the transform $\mathtt{M}$: the statement is broken exactly when
a warp consumes $\mathtt{M}$ with no earlier validate of $\mathtt{M}$.
\end{example}

\subject{Beyond the six forms.}
For a requirement that none of the forms expresses, a statement applies $\Box$ (``always'') directly
to a \emph{temporal body} $\Theta$: patterns combined with $\neg$, $\wedge$, $\Diamond^{-}$ (``at
some earlier point''), and $\Diamond^{+}$ (``at some later point''); the two diamonds are past- and
future-directed shorthands for the standard eventually operator $\Diamond$.
\appref{app:form-readings} writes out this reading for each preset form.
\begin{figure*}[t]
\centering
\footnotesize
\begin{mathpar}
\inferrule{\Sigma \vdash \varphi_i \rightsquigarrow \mathcal{R}_{\varphi_i} \quad (1 \le i \le m)}
          {\Sigma \vdash \{\varphi_1, \dots, \varphi_m\} \rightsquigarrow \textstyle\bigwedge_{i} \mathcal{R}_{\varphi_i}}{\ (\textsc{Policy})}

\inferrule{\mathcal{R}_\varphi \;=\; \textstyle\bigwedge_{e} \neg\,\psi(e)}
          {\Sigma \vdash {\sf abs}(\psi) \rightsquigarrow \mathcal{R}_\varphi}{\ (\textsc{Abs})}

\inferrule{\mathcal{R}_\varphi \;=\; \textstyle\bigwedge_{e} \Theta(e)}
          {\Sigma \vdash \Box\,\Theta \rightsquigarrow \mathcal{R}_\varphi}{\ (\textsc{Fallback})}

\inferrule{\mathcal{R}_\varphi \;=\; \textstyle\bigwedge_{e} \big(\psi(e) \Rightarrow \bigvee_{e' \prec e} \psi'(e')\big)}
          {\Sigma \vdash {\sf prec}(\psi, \psi') \rightsquigarrow \mathcal{R}_\varphi}{\ (\textsc{Prec})}

\inferrule{\mathcal{R}_\varphi \;=\; \textstyle\bigwedge_{e} \big(\psi(e) \Rightarrow \bigvee_{e' \succ e} \psi'(e')\big)}
          {\Sigma \vdash {\sf resp}(\psi, \psi') \rightsquigarrow \mathcal{R}_\varphi}{\ (\textsc{Resp})}

\inferrule{\mathcal{R}_\varphi \;=\; \textstyle\bigwedge_{e} \big(\psi(e) \Rightarrow \bigvee_{e \prec_\ell e'} \psi'(e')\big)}
          {\Sigma \vdash {\sf bresp}_\ell(\psi, \psi') \rightsquigarrow \mathcal{R}_\varphi}{\ (\textsc{BResp})}

\inferrule{\mathcal{R}_\varphi \;=\; \textstyle\bigwedge_{e} \big(\psi(e) \Rightarrow \psi'(e) \vee \bigvee_{e' \succ e} (\psi'(e') \vee \psi(e'))\big)}
          {\Sigma \vdash {\sf rslv}(\psi, \psi') \rightsquigarrow \mathcal{R}_\varphi}{\ (\textsc{Rslv})}

\inferrule{\mathcal{R}_\varphi \;=\; \textstyle\bigwedge_{e} \big(\psi(e) \Rightarrow \bigwedge_{e_b \succ e} (\psi_b(e_b) \Rightarrow \bigvee_{e \prec e_c \prec e_b} \psi'(e_c))\big)}
          {\Sigma \vdash {\sf until}(\psi, \psi', \psi_b) \rightsquigarrow \mathcal{R}_\varphi}{\ (\textsc{Until})}
\end{mathpar}
\smallskip
\hrule
\smallskip
\noindent and each pattern $\psi$ denotes a predicate $\psi(e)$ on one event:
\[
\begin{array}{r@{\;}c@{\;}l@{\hspace{1.6em}}r@{\;}c@{\;}l@{\hspace{1.6em}}r@{\;}c@{\;}l@{\hspace{1.6em}}r@{\;}c@{\;}l}
(a \in \alpha)(e) &\triangleq& a(e) \in \alpha & (k \in S)(e) &\triangleq& e[k] \in S & (k \sim p)(e) &\triangleq& e[k] \approx p & (s = \top)(e) &\triangleq& s(e) = \top \\[2pt]
(s = \bot)(e) &\triangleq& s(e) = \bot & (o \sim p)(e) &\triangleq& o(e) \approx p & (\psi \wedge \psi')(e) &\triangleq& \psi(e) \wedge \psi'(e) & & &
\end{array}
\]
and a temporal body $\Theta$ unfolds by the same finite quantification (a bare pattern $\psi$ is read
by its predicate above):
\[
\begin{array}{r@{\;}c@{\;}l@{\hspace{1.6em}}r@{\;}c@{\;}l@{\hspace{1.6em}}r@{\;}c@{\;}l@{\hspace{1.6em}}r@{\;}c@{\;}l}
(\neg\Theta)(e) &\triangleq& \neg\,\Theta(e) & (\Theta \wedge \Theta')(e) &\triangleq& \Theta(e) \wedge \Theta'(e) & (\Diamond^{-}\Theta)(e) &\triangleq& \textstyle\bigvee_{e' \prec e} \Theta(e') & (\Diamond^{+}\Theta)(e) &\triangleq& \textstyle\bigvee_{e' \succ e} \Theta(e')
\end{array}
\]
\caption{Symbolic compilation of the policy language, top down: a policy to the conjunction of its
statements (\textsc{Policy}), each statement to a requirement $\mathcal{R}_\varphi$ by one rule per
form (with \textsc{Fallback} for a bare temporal body), and each pattern and temporal-body operator
to a one-event predicate (foot).}
\label{fig:rules}
\end{figure*}

\subsection{Compiling Statements into Requirements}
\label{sec:rules}

A statement says what must hold; to test it against a trace, we need a formula the solver can check.
Compilation is syntax-directed, from a policy down to its patterns: \tool compiles each
statement $\varphi$ of the policy to its \emph{requirement} $\mathcal{R}_\varphi$ through the
judgment
\[
  \Sigma \vdash \varphi \rightsquigarrow \mathcal{R}_\varphi,
\]
read ``under signature $\Sigma$, statement $\varphi$ compiles to requirement $\mathcal{R}_\varphi$'',
also written ${\sf encode}(\varphi, \Sigma)$; \autoref{fig:rules} gives the rules.

The figure has three parts. At the top, \textsc{Policy} conjoins the statements' requirements: a
policy holds exactly when every statement does. The remaining rules build $\mathcal{R}_\varphi$, one
per form plus \textsc{Fallback}, and they share one shape: a guarded quantification over the events. The
outer conjunction ranges over all events, and for an event matching $\psi$ the body requires a
matching $\psi'$ at the right position in $\prec$ (earlier for ${\sf prec}$, later for ${\sf resp}$,
within the next $\ell$ events for ${\sf bresp}$, written $e \prec_\ell e'$, and so on). The foot
interprets the leaves: each pattern denotes a one-event predicate $\psi(e)$, and each temporal-body
operator unfolds into the same finite quantification.

A requirement is therefore one condition per event, conjoined; writing
$\mathcal{R}_\varphi(e)$ for the condition on event $e$,
\[
  \mathcal{R}_\varphi \;=\; \bigwedge_{e \in E} \mathcal{R}_\varphi(e).
\]
For ${\sf prec}$, $\mathcal{R}_\varphi(e)$ reads: if $e$ matches $\psi$, then some earlier event
matches $\psi'$; when the two patterns share the
variable $x$, each disjunct of a binary form gains the conjunct $e'[k'] = e[k]$, so only the
events that agree count. Every quantifier ranges over the finite event set $E$, and every set is drawn from the finite
signature, so each $\mathcal{R}_\varphi$ is a ground, quantifier-free formula over $\sem{\tau}$,
decided by an ordinary solver query.

\subsection{Deciding Requirements with Witnesses}
\label{sec:verify}

Checking a statement is a single solver query: the recorded facts either are consistent with the
requirement or refute it, and on a refutation the offending events are read off the solver's
unsatisfiable core.

\subject{Encoding.}
\tool asserts the fact base $\sem{\tau}$ together with the requirement, one condition
$\mathcal{R}_\varphi(e)$ per event, and asks whether the facts refute the requirement:
\[
  {\sf unsat}\big(\sem{\tau} \wedge \mathcal{R}_\varphi\big)
  \quad\Leftrightarrow\quad
  \tau \not\models \varphi .
\]
The equivalence holds because the fact base is complete: the facts fix everything the run did, so
they contradict $\mathcal{R}_\varphi$ exactly when the run breaks one of its conditions. Because
$\sem{\tau}$ pins down a finite structure and $\mathcal{R}_\varphi$ is quantifier-free over it, the
query is decidable. The checking problem thus reduces statement by statement,
\[
  \tau \not\models P \quad\Leftrightarrow\quad \bigvee_{\varphi \in P} {\sf unsat}\big(\sem{\tau} \wedge \mathcal{R}_\varphi\big),
\]
and the goal fixed by the problem statement is decided by $m$ such queries over one fact base: the
run satisfies $P$ exactly when no query is unsatisfiable.

\subject{Localization.}
An inconsistency establishes that the run is unsafe, but an operator acting on the verdict also
needs to know where it broke. The solver reports more than ${\sf unsat}$. ${\sf muc}$ returns a \emph{minimal unsatisfiable
core}: a smallest set of asserted conditions that already contradicts the facts. Each condition in the core names the
event it constrains, so localization is a lookup rather than a search:
\[
  \mathrm{loc}(\varphi) \;=\; \{\, e \in E \mid \mathcal{R}_\varphi(e) \in {\sf muc}(\sem{\tau} \wedge \mathcal{R}_\varphi) \,\},
\]
and one solver query yields the verdict and its core together.
For the binary forms, the core names an event matching $\psi$ whose required $\psi'$ is missing;
for ${\sf abs}$ it names an event exhibiting $\psi$, and for ${\sf until}$ it names the trigger
of the broken interval.

\begin{example}[Localizing the violation]
On the violating dashcam run, the warp step $e_5$ consumes the transform $\mathtt{M}$. The condition that R3's
requirement places on $e_5$ asks for an earlier validate of the value it consumes:
\[
\begin{aligned}
  \mathcal{R}_\varphi(e_5) \;=\; \psi_c(e_5) \Rightarrow{} & \\[-1pt]
  \bigvee_{e' \prec e_5} \big( \psi_v(e') \wedge e'[\icode{target}]& = e_5[\icode{transform}] \big).
\end{aligned}
\]
The facts make $\psi_c(e_5)$ true and fix $e_5[\icode{transform}] = \mathtt{M}$, but no event
before $e_5$ validates $\mathtt{M}$, so every disjunct is false: the facts and
$\mathcal{R}_\varphi(e_5)$ cannot hold together, and the query is unsatisfiable. Every other condition is consistent with the facts, so
${\sf muc}$ returns the singleton $\{\mathcal{R}_\varphi(e_5)\}$, and the lookup gives
$\mathrm{loc}(\varphi) = \{e_5\}$. \tool surfaces $e_5$ as the action to block.
\end{example}

\section{Implementation}
\label{sec:implementation}

We implement \tool in roughly 7,000 lines of Python~3.12 as a prototype of the framework in \autoref{fig:overview}.
The implementation uses Z3 as the backend SMT solver and \icode{claude-opus-4-6} for the specification-compilation step (\appref{app:prompt-template}), with \icode{fast-mode}.
The language model is used only to draft statements in the policy language; all grounding, validation, SMT encoding, and safe/unsafe decisions are performed by deterministic components.
Two implementation choices keep the language model contained.
First, in \textsc{Compile} the model may only use the vocabulary that \textsc{Abstract} recovers from the run, and a statement it proposes is accepted only after validation against the observed signature; clauses that cannot be grounded are reported as unsupported rather than silently kept.
Second, \textsc{Verify} invokes Z3 on each grounded statement independently and reports a witness per statement, so localization stays tied to the statement that failed.
For argument grounding, \textsc{Compile} can only reference parameter roles that \textsc{Abstract} binds to concrete trace events.
\tool grounds these roles from a small catalog of common POSIX and developer-tool commands, AST-based recovery for case-shipped Python, shell, and JavaScript scripts, and optional \icode{--help} probing for external CLIs.
\icode{--help} probing is opt-in because it runs the target binary.

\subject{Deployment.}
\tool runs outside the agent, in a sandboxed or mediated runtime that records the execution stream and can pause policy-relevant operations before their effects are released; this fits agent workflows that already mediate tool execution through a workspace, sandbox, or gateway.
At each enforcement point, the runtime presents \tool with the trace so far and the pending operation.
Prohibitions and preconditions are decided on the trace so far, since a witnessed violation cannot be undone; open obligations are judged when the run ends.

\subject{Enforcement actions.}
On a violating pending invocation, the runtime's pre-execution hook denies the call and returns the localized witness (Claude Code's pre-tool-use deny, or an MCP gateway holding a request); a blocked call has no effects to undo.
The deployment then halts the run, asks for user approval, or rolls the sandbox back to its last checkpoint and lets the agent replan: \tool decides whether the pending operation is safe, and the deployment chooses how to proceed.
In the real-world study (\autoref{sec:eval-rq4}), external CLIs are replaced by logging stubs, so verdicts gate the trace replay rather than live effects.

\section{Evaluation}
\label{sec:evaluation}

In this section, we present the design and results of the evaluation for answering the following research questions:
\begin{itemize}[noitemsep, topsep=1pt, leftmargin=1.5em]
  \item \textbf{RQ1 (Effectiveness).} How effectively does \tool detect specification violations in real LLM-agent trajectories without flagging benign workloads, compared with state-of-the-art runtime defenses?
  \item \textbf{RQ2 (Ablation).} What is the contribution of \tool's individual components?
  \item \textbf{RQ3 (Runtime Overhead).} What is the per-case cost of \tool, and how does it scale with the length of the trajectory?
  \item \textbf{RQ4 (Real-world Study).} Does \tool detect and block policy violations in real-world skill deployments?
\end{itemize}

\subsection{Experimental Setup}
\label{sec:eval-setup}

Our evaluation rests on a labeled corpus of real agent executions, each paired with its skill specification and a ground-truth violation label, against which we compare \tool with a set of runtime-enforcement baselines.
\begin{table*}[t]
\centering
\caption{Comparison of \tool with other policy-enforcement systems.}
\label{tab:rq1-overall}
\scriptsize
\setlength{\tabcolsep}{3.0pt}
\renewcommand{\arraystretch}{1.08}
\begin{tabular}{lc|ccccc|ccccc|ccccc|ccccc}
\toprule
\textbf{} & \textbf{} & \multicolumn{5}{c|}{\textbf{\tool}} & \multicolumn{5}{c|}{\textbf{AgentSpec}} & \multicolumn{5}{c|}{\textbf{Progent}} & \multicolumn{5}{c}{\textbf{LLM-as-judge}} \\
\midrule
	\textbf{Benchmark\textsuperscript{1}} & \textbf{Tests}
& TP & \cellcolor{black!5}FP & \cellcolor{black!10}Prec. & \cellcolor{black!15}Rec. & \cellcolor{black!20}F1
& TP & \cellcolor{black!5}FP & \cellcolor{black!10}Prec. & \cellcolor{black!15}Rec. & \cellcolor{black!20}F1
& TP & \cellcolor{black!5}FP & \cellcolor{black!10}Prec. & \cellcolor{black!15}Rec. & \cellcolor{black!20}F1
& TP & \cellcolor{black!5}FP & \cellcolor{black!10}Prec. & \cellcolor{black!15}Rec. & \cellcolor{black!20}F1 \\
\midrule
\textsc{SB+SI} & 152
& 69 & \cellcolor{black!5}8 & \cellcolor{black!10}89.6\% & \cellcolor{black!15}95.8\% & \cellcolor{black!20}92.6\%
& 34 & \cellcolor{black!5}12 & \cellcolor{black!10}73.9\% & \cellcolor{black!15}47.2\% & \cellcolor{black!20}57.6\%
& 40 & \cellcolor{black!5}47 & \cellcolor{black!10}46.0\% & \cellcolor{black!15}55.6\% & \cellcolor{black!20}50.3\%
& 48 & \cellcolor{black!5}5 & \cellcolor{black!10}90.6\% & \cellcolor{black!15}66.7\% & \cellcolor{black!20}76.8\% \\
AgentDojo~\cite{debenedetti2024agentdojo} & 629
& 298 & \cellcolor{black!5}35 & \cellcolor{black!10}89.5\% & \cellcolor{black!15}99.3\% & \cellcolor{black!20}94.2\%
& 290 & \cellcolor{black!5}41 & \cellcolor{black!10}87.6\% & \cellcolor{black!15}96.7\% & \cellcolor{black!20}91.9\%
& 285 & \cellcolor{black!5}49 & \cellcolor{black!10}85.3\% & \cellcolor{black!15}95.0\% & \cellcolor{black!20}89.9\%
& 297 & \cellcolor{black!5}34 & \cellcolor{black!10}89.7\% & \cellcolor{black!15}99.0\% & \cellcolor{black!20}94.1\% \\
SafeAgentBench~\cite{yin2024safeagentbench} & 546
& 234 & \cellcolor{black!5}4 & \cellcolor{black!10}98.3\% & \cellcolor{black!15}94.4\% & \cellcolor{black!20}96.3\%
& 231 & \cellcolor{black!5}2 & \cellcolor{black!10}99.1\% & \cellcolor{black!15}93.1\% & \cellcolor{black!20}96.0\%
& \multicolumn{5}{c|}{N/A}
& 222 & \cellcolor{black!5}41 & \cellcolor{black!10}84.4\% & \cellcolor{black!15}89.5\% & \cellcolor{black!20}86.9\% \\
\bottomrule
\end{tabular}
	\vspace{0.5mm}
	\parbox{\textwidth}{\raggedright\footnotesize
	\textsuperscript{1}AgentDojo and SafeAgentBench rows reuse cases, not native metrics. We score TP, FP, precision, recall, and F1 under our policy-enforcement protocol. Progent is N/A on SafeAgentBench: its release targets tool-call permissions and lacks an embodied-action adapter.}
	\end{table*}

\subject{Benchmark.}
We construct the main evaluation set from SkillsBench~\cite{li2026skillsbench} and Skill-Inject~\cite{schmotz2026skill}, combining them into a final labeled evaluation set, denoted \textsc{SB+SI}. This set contains 152 executions: 72 policy-violating trajectories and 80 benign trajectories. We use AgentDojo~\cite{debenedetti2024agentdojo} and SafeAgentBench~\cite{yin2024safeagentbench} only for additional cross-benchmark comparisons. Further details on labeling are in \appref{app:labeling-protocol}.
\begin{itemize}[leftmargin=1.5em, itemsep=2pt, topsep=4pt]
    \item \textbf{SkillsBench.} SkillsBench provides over 4,000 task-performance trajectories. We use an LLM to screen for trajectories whose skill documentation states an enforceable behavioral policy and whose execution appears relevant to that policy. We then manually inspect each candidate to label whether it violates or follows the policy.
    \item \textbf{Skill-Inject.} It does not provide raw execution trajectories, we replay its tasks in Claude Code using \icode{claude-sonnet-4-6} with \icode{bypassPermissions}, then record the resulting agent executions. From these replayed executions, we identify and label trajectories that violate the behavioral policy stated by the skill and trajectories that do not.
    \item \textbf{Additional comparisons.} AgentDojo is a dynamic benchmark for prompt-injection attacks and defenses in tool-using LLM agents, and SafeAgentBench evaluates safe task planning in embodied LLM agents. We use both as cross-benchmark comparisons under the settings used by the corresponding baselines. For AgentDojo, we use the published trajectories generated by \icode{gpt-4o-2024-05-13}.
\end{itemize}

\subject{Baselines.}
We select three direct baselines to represent the prevailing paradigms in agent policy enforcement:
\begin{itemize}[leftmargin=1.5em, itemsep=2pt, topsep=4pt]
    \item \textbf{AgentSpec}~\cite{wang2025agentspec} represents \emph{action-boundary enforcement}. Its rule language can state temporal triggers, but each rule is evaluated when its trigger action is proposed, so the decision sees one pending action at a time and cannot reach evidence that lives elsewhere in the trace, such as artifact identity or long-range event order.
    \item \textbf{Progent}~\cite{shi2025progent} similarly represents \emph{action-boundary guardrails}, compiling policies into per-tool-call permission allowlists.
    \item \textbf{LLM-as-judge} uses \icode{claude-opus-4-6} as a model-only whole-trace judge. It observes the full trace context, but operates without deterministic formal semantics and cannot produce a reliable, localized witness for runtime blocking.
\end{itemize}

Together, these baselines allow us to compare \tool's \emph{trace-grounded} approach against both localized and informal enforcement strategies. For AgentSpec and Progent, we count a blocked action as a reported violation; for LLM-as-judge, we count a positive judgment as a reported violation (\appref{app:baseline-adaptation}).

\subject{Metrics.}
For violation detection, we report TP, FP, precision, recall, and F1 per system; any reported violation on a benign trace counts as a false positive.
For overhead, we report end-to-end latency and stage-level latency.

\subject{Experiment settings.}
We ran all experiments on a server running Ubuntu 22.04.5 LTS with 16 logical Intel Xeon Gold 5222 CPUs and 125 GiB of RAM.

\subsection{RQ1. Effectiveness}
\label{sec:eval-rq1}
We evaluate \tool, AgentSpec, Progent, and the LLM-as-judge baseline on \textsc{SB+SI}.
Each system receives the same specification-derived policy and execution trace, and we score whether it reports a policy violation under its intended enforcement interface.
Representative policy categories and formalizations for \textsc{SB+SI} appear in \appref{app:policy-forms}.

\subject{Overall results.}
\autoref{tab:rq1-overall} compares \tool with the baselines and shows a stark contrast between benchmarks that require trace-level reasoning and those decided by single-call.

On our primary target, \textsc{SB+SI}, where violations turn on temporal structure and value binding, \tool detects 69 of the 72 violating executions with only 8 false positives on 80 benign runs (92.6\% F1). Action-boundary defenses cannot track historical artifacts or prerequisites: AgentSpec and Progent miss about half the violations (57.6\% and 50.3\% F1). The LLM-as-judge baseline sees the full context and does better (76.8\% F1), but still misses 24 temporal violations for lack of formal grounding. \tool leads the strongest baseline by 15.8 F1 points.

The AgentDojo and SafeAgentBench rows serve as cross-benchmark sanity checks. Their labeled violations are largely visible at a single action boundary (e.g., a prohibited command string), exactly what AgentSpec and Progent are optimized for. AgentDojo also contains semantic prompt-injection conflicts that a strong whole-trace judge such as \icode{claude-opus-4-6} can often classify directly. Even so, \tool holds its own without per-benchmark tuning: it is the strongest system on both SafeAgentBench (96.3\% vs 96.0\% F1) and AgentDojo (94.2\% vs 94.1\% F1), albeit by narrow margins. \textsc{SB+SI} remains the setting that measures the trace-level violations \tool is built for.

These results reflect enforcement granularity. The same \icode{bash} call may be benign or violating depending on the file it targets, an earlier validation, and the artifact it leaves behind; \tool improves recall by deciding over grounded events across the full trace rather than at the pending action alone.

\subject{Measuring enforcement reliability.}
A runtime monitor fails in exactly two ways: it misses a real violation, or it falsely blocks a benign run. Either error can be hidden by accepting more of the other: flag everything and recall is perfect, flag nothing and false alarms vanish. \autoref{fig:rq1-bee} therefore reports the one number that punishes both, the \emph{balanced enforcement error} (BEE):
\[
  \mathrm{BEE} \;=\; \tfrac{1}{2}\,\big(\mathrm{FNR} + \mathrm{FPR}\big),
\]
where the $\mathrm{FNR}$ is the fraction of violating executions a system misses and the $\mathrm{FPR}$ is the fraction of benign executions flagged; BEE is the complement of balanced accuracy. Thus, BEE is how often the monitor gets the call wrong, counting a missed violation and a blocked legitimate run as equally costly. We macro-average over the three benchmark sets so each counts equally. \tool attains the lowest macro BEE (5.4\%), less than half that of the next best complete baseline: reducing both error types rather than optimizing recall alone.

\begin{figure}[t]
\centering
\begin{tikzpicture}
\begin{axis}[
  xbar,
  width=0.95\columnwidth, height=2.7cm,
  bar width=8pt,
  xmin=0, xmax=17.5,
  ymin=-0.55, ymax=2.55,
  xlabel={Macro balanced enforcement error (\%), lower is better},
  xlabel style={font=\scriptsize\sffamily},
  ytick={0,1,2},
  yticklabels={AgentSpec, LLM-as-judge, \tool},
  yticklabel style={font=\scriptsize\sffamily},
  xticklabel style={font=\scriptsize\sffamily},
  xmajorgrids, major grid style={black!9},
  axis x line*=bottom, axis y line*=left,
  axis line style={black!85},
  tick style={black!85},
  nodes near coords,
  every node near coord/.append style={font=\fontsize{5.5}{6}\selectfont\sffamily,
                                       text=tgInk, anchor=west},
  nodes near coords style={/pgf/number format/.cd, fixed, fixed zerofill, precision=1},
]
\addplot[xbar, fill=black!8, draw=black!60, line width=0.3pt, bar shift=0pt,
         postaction={pattern=north east lines, pattern color=black!40}]
  coordinates {(15.2,0)};
\addplot[xbar, fill=black!20, draw=black!60, line width=0.3pt, bar shift=0pt,
         postaction={pattern=horizontal lines, pattern color=black!45}]
  coordinates {(12.5,1)};
\addplot[xbar, fill=black!55, draw=black!75, line width=0.3pt, bar shift=0pt]
  coordinates {(5.4,2)};
\end{axis}
\end{tikzpicture}
\caption{Cross-benchmark balanced enforcement error, macro-averaged over the
three benchmark sets. Progent appears only in \autoref{tab:rq1-overall}, as it
does not run on SafeAgentBench.}
\label{fig:rq1-bee}
\end{figure}
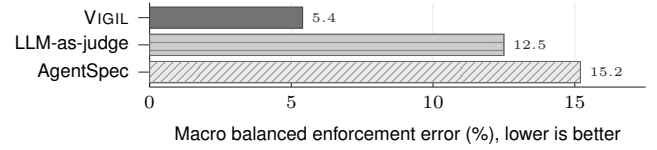
\begin{figure*}[t]
\centering
\begin{minipage}[t]{0.48\textwidth}
\centering
  \begin{tikzpicture}[
    rec/.style={draw=black!65, rounded corners=2pt, inner sep=3pt,
                font=\scriptsize, align=center, text width=4.6cm,
                fill=recfill},
    inner/.style={draw=black!50, dashed, rounded corners=2pt, inner sep=3pt,
                  font=\scriptsize, align=center, text width=4.6cm, fill=black!4},
    A/.style={-{Stealth[length=1.6mm, width=1.4mm]}, line width=0.9pt},
    Ad/.style={-{Stealth[length=1.6mm, width=1.4mm]}, line width=0.9pt,
               dashed, draw=black!55},
    BL/.style={font=\scriptsize\itshape, text=black!70, align=left,
               text width=2.5cm, anchor=west},
    Tag/.style={font=\scriptsize\itshape, text=black!65, align=left,
                anchor=west},
  ]
    \node[rec] (call) at (0, 0)
      {agent invokes\\\icode{secure\_archive.py numbers.xlsx}};
    \node[Tag] at ([xshift=6pt]call.east) {recorded as~$e_{26}$};

    \node[inner, below=0.22cm of call] (s1)
      {open ZIP (no real encryption)};
    \node[inner, below=0.18cm of s1] (s2)
      {\icode{zf.setpassword} (no-op on write)};
    \node[inner, below=0.18cm of s2] (s3)
      {\icode{input\_file.unlink()}};

    \node[rec, below=0.22cm of s3] (return)
      {control returns to agent \textit{(stdout only)}};
    \node[Tag] at ([xshift=6pt]return.east) {recorded as~$e_{27}$};

    \draw[A]  (call) -- (s1);
    \draw[Ad] (s1)   -- (s2);
    \draw[Ad] (s2)   -- (s3);
    \draw[A]  (s3)   -- (return);
    \draw[decorate, decoration={brace, amplitude=4pt, raise=4pt},
          line width=0.7pt, draw=black!55]
      (s1.north east) -- (s3.south east);
    \node[BL] at ([xshift=14pt]$(s1.east)!0.5!(s3.east)$)
      {no trace facts recorded for these steps\\(they live inside the script body)};
  \end{tikzpicture}
\par\smallskip
(a) False negative: \icode{secure\_archive.py}
\end{minipage}\hfill
\begin{minipage}[t]{0.48\textwidth}
\centering
  \begin{tikzpicture}[
    rec/.style={draw=black!65, rounded corners=2pt, inner sep=3pt,
                font=\scriptsize, align=left, text width=5.6cm,
                fill=recfill},
    inner/.style={draw=black!50, dashed, rounded corners=2pt, inner sep=3pt,
                  font=\scriptsize\ttfamily, align=left, text width=5.6cm,
                  fill=black!4},
    A/.style={-{Stealth[length=1.6mm, width=1.4mm]}, line width=0.9pt},
    BL/.style={font=\scriptsize\itshape, text=black!70, align=left,
               text width=2.1cm, anchor=west},
    Tag/.style={font=\scriptsize\itshape, text=black!65, align=left,
                anchor=west},
  ]
    \node[rec] (policy) at (0, 0)
      {``\textbf{Bulk} modifications to calendar data without
        explicit user confirmation risk loss of important
        scheduling information''};
    \node[Tag] at ([xshift=6pt]policy.east) {policy};

    \node[inner, below=0.28cm of policy] (compiled)
      {${\sf abs}$($a \in \{\icode{Edit}\} \wedge \icode{file\_path} \sim \icode{"*calendar*"}$)};

    \node[rec, below=0.28cm of compiled] (event)
      {\ttfamily Edit(file\_path=".../calendar.ics",\\
       \hspace*{1.6em}old\_string="...090000Z",\\
       \hspace*{1.6em}new\_string="...120000Z")};
    \node[Tag] at ([xshift=6pt]event.east) {event};

    \draw[A] (policy)   -- (compiled);
    \draw[A] (compiled) -- (event);

    \draw[decorate, decoration={brace, amplitude=4pt, raise=4pt},
          line width=0.7pt, draw=black!55]
      (compiled.north east) -- (compiled.south east);
    \node[BL] at ([xshift=8pt]$(compiled.east)$)
      {the policy statement drops the
        qualifier ``\textbf{bulk}''};
  \end{tikzpicture}
\par\smallskip
(b) False positive: benign calendar edit
\end{minipage}
\caption{\tool's two error boundaries. (a) A false negative at the boundary of trace
observability: the policy-relevant steps run inside the script body and produce no trace
facts. (b) A false positive from compilation fidelity: the generated statement drops the
qualifier \emph{bulk} and fires on a benign single-event edit.}
\label{fig:eval-error-cases}
\end{figure*}
\subject{Error analysis.}
The baselines err in different directions. Progent over-blocks benign executions because it turns the policy into a narrow allowlist: \textsc{SB+SI} policies rarely enumerate every safe command, path, or argument, and all 47 of its false positives are benign calls the allowlist rejects. The LLM-as-judge errs the other way, missing 24 violations. \tool avoids both by checking grounded trace events rather than a complete allowlist or an unconstrained model verdict.

\tool's own remaining errors mark two system boundaries. False negatives mark the \emph{boundary of trace observability}: policy-relevant evidence can stay hidden inside an opaque utility script. In \autoref{fig:eval-error-cases}a, the agent invokes \icode{secure\_archive.py}; the script creates an unencrypted ZIP (a no-op \icode{zf.setpassword}) and immediately \icode{unlink}s the source file. The policy requires an integrity check between write and deletion, but both actions run inside the script body and never surface as trace events.

False positives mark the \emph{boundary of compilation fidelity}: the LLM can translate a specification into an overly broad statement or a wrong temporal form. In \autoref{fig:eval-error-cases}b, the specification limits a rule to \emph{bulk} calendar modifications, the generated statement drops the qualifier, and the monitor fires on a benign single-event reschedule. In these cases, the error is semantic over-generalization, not a failure to compile. Both boundaries define clear future work: deeper sandbox instrumentation, and stricter compilation prompting.

\takeaways{\tool detects 69 of 72 real violations at 95.8\% recall and 89.6\% precision, leading the strongest state-of-the-art baseline by 15.8 F1 points. \tool sets a new state of the art for detecting behavioral violations in real agent executions.}

\subsection{RQ2. Ablation Study}
\label{sec:eval-rq2}

We ablate the two structures that make operation understanding available before solver checking.
All variants use the same \textsc{SB+SI} evaluation set, labels, and metrics as RQ1, and each removes one source of structure while leaving the rest of the pipeline unchanged:
\begin{enumerate}[leftmargin=1.5em, itemsep=2pt, topsep=4pt]
    \item \textbf{w/o Argument Grounding} removes the grounding of command and script arguments into event fields, so the monitor still observes tool calls but loses which values within an event can be referenced by the policy.
    \item \textbf{w/o Policy Structure} removes the structured intermediate form used when translating natural-language specifications into executable policies, testing whether the compilation step preserves ordering constraints, object identities, and final-state obligations without that.
\end{enumerate}

\begin{table}[t]
\centering
\caption{Ablation results on \textsc{SB+SI} (152 cases).}
\label{tab:rq2-ablation}
\footnotesize
\setlength{\tabcolsep}{2.5pt}
\renewcommand{\arraystretch}{1.08}
\begin{tabular}{lrrrrrr}
\toprule
\textbf{Configuration} & \textbf{Prec.} & \textbf{Rec.} & \textbf{FPR} & \textbf{F1} & \textbf{$\Delta$Prec.} & \textbf{$\Delta$Rec.} \\
\midrule
\rowcolor{tgTableRate} \tool & \textbf{89.6\%} & \textbf{95.8\%} & \textbf{10.0\%} & \textbf{92.6\%} & --- & --- \\
w/o Arg.\ Grounding & 78.0\% & 63.9\% & 16.2\% & 70.2\% & \textcolor{black!70}{$\downarrow$}\,11.6 & \textcolor{black!70}{$\downarrow$}\,31.9 \\
w/o Policy Struct. & 86.7\% & 90.3\% & 12.5\% & 88.4\% & \textcolor{black!70}{$\downarrow$}\,2.9 & \textcolor{black!70}{$\downarrow$}\,5.5 \\
\bottomrule
\end{tabular}
\end{table}

\autoref{tab:rq2-ablation} reports the ablation results.
Removing argument grounding reduces precision by 11.6 percentage points and recall by 31.9 points.
Without grounded arguments, \textsc{Compile} cannot connect conditions to the values that matter within an event, so the monitor falls back to action-only matches that miss most argument-dependent violations.
The missed cases are concentrated in protocols whose violation depends on the meaning of command arguments, such as paths, modes, script targets, payload strings, or target objects.
Removing policy structure reduces precision by 2.9 percentage points and recall by 5.5 points.
The compiler still produces executable statements, but more often emits broad single-event ${\sf abs}$ statements that miss temporal/final-state violations or fire on benign payloads that match the generated condition.
These misses are concentrated in protocols whose natural-language form is not a simple forbidden action, including ordering constraints, prerequisite checks, and final-artifact requirements.
Together, the ablations show that \tool depends on structured evidence before solver checking: grounded event values and policy structure.

\takeaways{Removing argument grounding costs 31.9 recall points; removing policy structure costs another 5.5. Both components contribute decisively: together they provide the operation understanding that powers \tool's leading accuracy.}
\begin{table*}[t]
\centering
\caption{Representative real-world findings across deployed skill ecosystems.}
\label{tab:rq4-cases}
\footnotesize
\setlength{\tabcolsep}{3.3pt}
\renewcommand{\arraystretch}{1.2}
\begin{tabular}{@{}l
                >{\raggedright\arraybackslash}p{0.385\textwidth}
                >{\raggedright\arraybackslash}p{0.285\textwidth}
                >{\raggedright\arraybackslash}p{0.115\textwidth}@{}}
\toprule
\textbf{Ecosystem} & \textbf{Root cause} & \textbf{Observed violation} & \textbf{Impact} \\
\midrule
\rowcolor{black!5}
Trail of Bits~\cite{trailofbits2025skills} &
\textbf{Specification gap}: ASan/fuzzer obligation split over skills &
ASan fuzzing with memory limits enabled &
Missed crashes \\
NVIDIA~\cite{nvidia2026slurmskills} &
\textbf{Specification gap}: Slurm workflow lacks a retry bound &
Unbounded Slurm GPU resubmissions &
GPU cost \\
\rowcolor{black!5}
Cloudflare~\cite{cloudflare2026skills} &
\textbf{Specification conflict}: example violates bundle secret rule &
Secret passed through shell into \icode{wrangler} &
Secret exposure \\
Databricks~\cite{databricks2025skills} &
\textbf{Agent execution gap}: CLI-version gate not propagated &
Commands ran without CLI-version gate &
Partial deploy \\
\rowcolor{black!5}
Anthropic ~\cite{anthropic2026xlsxskill} &
\textbf{Agent execution gap}: ignored \icode{data\_only} save warning &
Workbook saved after \icode{data\_only=True} &
Formula loss \\
Microsoft Deep Wiki~\cite{microsoft2026deepwiki} &
\textbf{Agent execution gap}: ignored ``never overwrite'' rule &
Existing \icode{AGENTS.md} overwritten &
Context loss \\
\bottomrule
\end{tabular}
\end{table*}

\subsection{RQ3. Runtime Overhead}
\label{sec:eval-rq3}

We measure \tool's overhead on all \textsc{SB+SI} traces and decompose
end-to-end latency into the three stages in \autoref{fig:overview}.
\autoref{tab:rq3-runtime-breakdown} shows that \tool averages 4.38\,s per
trace, with \textsc{Compile}'s single LLM call accounting for 92.5\%.

\begin{table}[t]
\centering
\caption{Runtime breakdown on \textsc{SB+SI}.}
\label{tab:rq3-runtime-breakdown}
\footnotesize
\setlength{\tabcolsep}{4.2pt}
\renewcommand{\arraystretch}{1.08}
\begin{tabular}{rrrr}
\toprule
\textbf{\textsc{Abstract}} & \textbf{\textsc{Compile}} & \textbf{\textsc{Verify}} & \textbf{End-to-end} \\
\midrule
0.063\,s (1.4\%) & 4.05\,s (92.5\%) & 0.27\,s (6.2\%) & \textbf{4.38\,s} \\
\bottomrule
\end{tabular}
\end{table}

\begin{figure}[t]
  \centering
  \includegraphics[width=0.8\columnwidth]{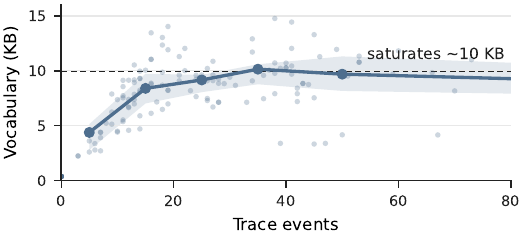}
  \caption{The vocabulary exposed to \textsc{Compile} saturates, so \tool's
  overhead stays stable as traces grow.}
  \label{fig:rq3-vocab-saturation}
\end{figure}

This cost does not grow with trace length. \textsc{Compile}'s input is the
observed vocabulary of the run, and \autoref{fig:rq3-vocab-saturation} shows it
saturating near 10\,KB once the initial events have introduced the run's tools,
argument names, and artifact types; end-to-end runtime therefore stays stable
across the full range of trace lengths.

For modern AI coding-agent workloads, where standard Bash commands average 4--6\,s and tasks run for several minutes~\cite{zheng2026agentcgroup}, a 4-second overhead is well within acceptable limits. More importantly, because the bottleneck is entirely within the LLM translation step rather than the SMT solver, it is highly optimizable. Deployments can cache compiled policies per skill, utilize faster model serving, or deploy small local distilled models to reduce end-to-end latency, while the determinism of the sub-second \textsc{Verify} path remains uncompromised.

\takeaways{The online check decides in 0.27\,s; 92.5\% of the 4.38\,s end-to-end cost is one cacheable, optimizable LLM compilation call, and the total plateaus as the trace vocabulary saturates. At agent timescales, \tool's enforcement is effectively free: a fraction of a single tool call buys a solver-backed verdict on every action.}

\subsection{RQ4. Real-world Skill-Bundle Study}
\label{sec:eval-rq4}

To assess \tool in the real world, we evaluate it on skill bundles drawn from deployed ecosystems. We deployed \tool over 216 executions sampled across three tiers: default-enabled harness bundles (e.g., Claude Code, Codex); first-party vendor ecosystems (e.g., NVIDIA, Databricks); and widely-used open-source catalogs\footnote{Marketplace/catalog bundles are assembled from public marketplaces~\cite{clawhub2026} where users install multiple third-party skills together.}.

Across these monitored executions, \tool surfaced 34 confirmed, real-world policy-violating runs out of 44 alerts. These violations include both skill-attributed defects, such as missing or conflicting bundle specifications, and agent-attributed failures, where the agent does not preserve or enforce a stated skill rule at runtime. We have a responsibility to disclose skill-attributed findings to affected maintainers through available channels. For agent-attributed failures, the skill specification already states the rule; the issue is that the agent runtime does not enforce it. One skill-attributed issue has been acknowledged by NVIDIA.

For each bundle, we used the BenchFlow SDK~\cite{benchflow2026} to mount the skill documents in a Docker container and replaced external CLIs with logging stubs. In this replay-safe study, \tool identifies the call it would block in a live deployment and outputs the offending witness before the stubbed effect is released. The 34 confirmed violations fall into three recurring patterns:
\begin{enumerate}[leftmargin=1.5em, itemsep=2pt, topsep=4pt]
    \item composition-level obligations missing across otherwise coherent skills;
    \item persistent policy obligations lost across a task-driven trace;
    \item stated skill rules not operationalized during execution.
\end{enumerate}
\autoref{tab:rq4-cases} separates failure origin: the first three findings expose \emph{specification gaps}, where skill documents leave obligations underspecified or conflicting, while the last three expose \emph{agent execution gaps}, where specifications state rules but agents fail to enforce them. We examine two in depth.

\subject{Case study 1: composition-level underspecification (Trail of Bits / NVIDIA).}
The first case concerns a 15-skill fuzzing bundle drawn from
Trail of Bits' open-source testing-handbook skills~\cite{trailofbits2025skills}.
The bundle combines \icode{address-sanitizer}, \icode{aflpp}, and
\icode{libfuzzer}. The relevant skill documents state that AddressSanitizer
maps, but does not reserve, 16+\,TB of virtual address
space~\cite{serebryany2012asan}, and instruct agents to disable the
corresponding fuzzer memory limit when ASan is enabled:
\icode{-m~none} for AFL++ and \icode{-rss\_limit\_mb=0} for libFuzzer.
These requirements appear under different flag names and mechanisms
(\icode{RLIMIT\_AS} for AFL++ and an RSS watchdog for libFuzzer), and the
skill documents do not state the cross-skill obligation explicitly. In the
probed trace, the agent compiled with \icode{-fsanitize=address} but invoked
both fuzzers without the required memory-limit-disabling flags. This
configuration can silently invalidate the campaign.
Under a global policy derived from the LLVM AddressSanitizer warning on
memory-limiting tools~\cite{llvm-asan-docs}, any fuzzer invocation after an
ASan-instrumented compilation must explicitly disable the corresponding
fuzzer memory limit.
\tool localizes the violation to the misconfigured fuzzer invocation.
We reported a similar composition-level issue for NVIDIA's skills, where a
missing liveness specification caused repeated Slurm GPU
submissions after evaluation failures; NVIDIA acknowledged this as a real cost concern.

\subject{Case study 2: missed prerequisite in execution (Databricks).}
The second case concerns a precondition violation in a Databricks skill
bundle~\cite{databricks2025skills}. The \icode{databricks-core} skill states
an explicit precondition: \emph{``If the CLI is missing or outdated
(\textless\,v0.292.0), stop immediately and do not attempt workarounds.''}
However, operational skills such as \icode{databricks-dabs} and
\icode{databricks-pipelines} instruct the agent to run commands like
\icode{databricks bundle deploy} without referencing this version gate. In the
probed trace, the agent's version check returned a non-matching result, yet
the agent proceeded to execute \icode{databricks} commands. The applicable
policy is a \emph{Precedence} property: every \icode{databricks} command must be
preceded by a successful version check confirming CLI $\geq$\,0.292.
The probed trace contains no such event, so \tool catches the gap and blocks the deployment.
The gate exists precisely because operational commands can misbehave on an outdated CLI; an agent that skips it can leave a bundle deployment half-applied.

\takeaways{On 216 executions of shipped skill bundles, \tool surfaces 34 confirmed policy-violating runs and localizes each to the offending call. The findings span NVIDIA, Databricks, Cloudflare, and Trail of Bits, covering both skill-specification defects and agent runtime enforcement gaps; NVIDIA acknowledged one skill-attributed issue.}

\section{Related Work}

\tool builds on runtime enforcement for AI agents, authorization policy languages, and agent-skill security, but differs by grounding each policy in the observed trace.

\subject{Runtime enforcement for AI agents.}
Runtime enforcement moves AI-agent safety beyond prompt-level guidance to mechanisms that monitor tool use, constrain data influence, and enforce policies~\cite{kim2026attack,kim2026sok}.
Action-boundary systems such as AgentSpec~\cite{wang2025agentspec}, Progent~\cite{shi2025progent}, Conseca~\cite{tsai2025contextual}, and PCAS~\cite{palumbo2026policy} guard proposed tool calls or compile contextual policies into instrumented checks.
Context- and data-aware systems such as CaMeL~\cite{debenedetti2025defeating}, IFC-based agents~\cite{costa2025securing,cai2026ghost}, AgentArmor~\cite{wang2025agentarmor}, and AgentSentry~\cite{zhang2026agentsentry} constrain actions using isolation, provenance, program analysis, or temporal causal diagnostics.
Together, they decide whether proposed actions proceed, are sanitized, or are constrained by policy, provenance, isolation, or data flow.

Execution-level work~\cite{ruan2024identifying,pysklo2026agent,wu2026policy} treats trajectories, effects, and world-state updates as the object of analysis.
However, they use executions mainly as objects to classify or validate, rather than as contexts that shape enforcement.
\tool instead grounds each behavioral specification in the observed trace, refining the policy to the concrete run and capturing relational or temporal violations in executions.

\subject{Policy languages.}
Policy languages express security rules over users, actions, resources, and conditions while leaving enforcement to a separate monitor or decision procedure~\cite{blaze1999keynote,ligatti2005edit,damianou2001ponder}.
Cedar~\cite{cutler2024cedar} provides a domain-specific language for fine-grained authorization policies, and major cloud platforms such as AWS IAM~\cite{awsIAMPolicies}, Azure RBAC~\cite{azureRBAC}, and Google Cloud IAM~\cite{googleCloudIAM} provide established policy languages for cloud resources.
Zelkova~\cite{backes2018zelkova} further shows that IAM-style policies can be given formal semantics and checked with SMT.
However, these languages operate at the resource-authorization level: they decide whether a principal may perform an action on a resource under a set of attributes.
Agent skill policies require a lower-level execution view: their policy object is a finite run with tool outputs, argument relations, and temporal ordering across calls, not a single authorization request.

Recent agent policy languages adapt policy ideas to agent actions: AgentSpec~\cite{wang2025agentspec} and Progent~\cite{shi2025progent} express action-level rules, while Conseca~\cite{tsai2025contextual} and PCAS~\cite{palumbo2026policy} generate or compile contextual policies for agent tasks.
However, existing agent policy languages typically tie policies to a relatively fixed action or task context.
\tool instead treats the observed trace as the policy context, so the enforceable specification can be refined online as the run unfolds.

\subject{Agent skill security.}
Agent skills package procedural knowledge, instructions, executable code, and resources into reusable capabilities; this expands the agent security boundary from model responses and tool APIs to the skill package itself~\cite{xu2026agent,jiang2026sok}.
Recent studies characterize skill architectures and lifecycle risks~\cite{xu2026agent,jiang2026sok,li2026towards}, and show that public skill ecosystems contain vulnerable, malicious, and harmful skills~\cite{liu2026agent,liu2026malicious,jiang2026harmfulskillbench}.
Recent systems audit or benchmark these risks: Skill-Inject~\cite{schmotz2026skill} studies skill-file injection, AgentTrap~\cite{zhuang2026agenttrap} evaluates runtime trust failures, SkillProbe~\cite{guo2026skillprobe} audits marketplace skills, and Semia~\cite{wen2026semia} converts skill text and structure into Datalog for static analysis.
Closest to our setting, Sefz~\cite{li2026no} shows that benign inputs can trigger specification violations when a skill breaks its own natural-language guardrails.
\tool complements this line by turning such behavioral requirements into runtime enforcement.
Instead of searching for inputs that expose a violation, \tool grounds the specification in the observed trace and checks whether the concrete run satisfies it.

\section{Conclusion}\label{sec:conclusion}

We presented \tool, a runtime reference monitor that makes agent-skill
behavioral specifications executable. \tool grounds each specification in the
agent's own execution trace and checks it with SMT, returning a localized
witness when a run violates the policy. On real LLM-agent runs, \tool detects
trace-level violations that single-call defenses miss (95.8\% recall at 89.6\%
precision) and surfaces 34 confirmed violations across deployed skill
ecosystems, including one specification defect acknowledged by NVIDIA. Agent
skills are still trusted to keep the contracts they declare; \tool makes those
contracts enforceable at runtime.

\bibliographystyle{IEEEtran}
\bibliography{references,agentsec,llm,policy_language,skill}

\newpage
\appendices

\makeatletter
\twocolumn[
\begin{@twocolumnfalse}
\begin{center}
\captionof{table}{Representative policy categories and formalizations in \textsc{SB+SI}.}
\label{tab:rq1-policy-forms}
\small
\setlength{\tabcolsep}{5.5pt}
\renewcommand{\arraystretch}{1.12}
\begin{tabular}{p{0.18\textwidth}p{0.34\textwidth}>{\footnotesize}p{0.40\textwidth}}
\toprule
\textbf{Category} & \textbf{Example policy} & \textbf{Formalization} \\
\midrule
\rowcolor{black!5}
Required event property &
Period search must include flux uncertainties. &
\(\Box(\mathtt{periodSearch}(x) \Rightarrow \mathtt{hasFluxErr}(x))\) \\
Authorization constraint &
Transfers require approved destinations. &
\(\Box(\mathtt{transfer}(x,d) \Rightarrow \mathtt{approvedDest}(d))\) \\
\rowcolor{black!5}
Temporal precondition &
Validate archives before source removal. &
\(\Box(\mathtt{removeSource}(x) \Rightarrow \Diamond^{-}\mathtt{validArchive}(y,x))\) \\
Final-state integrity &
Delivered Excel models must have zero formula errors. &
\(\Box(\mathtt{deliver}(x) \Rightarrow \mathtt{noFormulaErr}(x))\) \\
\bottomrule
\end{tabular}
\end{center}
\vspace{0.6em}
\end{@twocolumnfalse}
]
\makeatother

\section{Labeling Protocol for \textsc{SB+SI}}
\label{app:labeling-protocol}

We construct \textsc{SB+SI} by combining violation-oriented candidate mining with randomly sampled benign executions.
For SkillsBench, we first screen the released task-performance trajectories for cases in which the skill documentation states a behavioral obligation that can be checked against an execution trace, such as an ordering requirement, prerequisite check, argument constraint, or artifact-use rule.
This screen is used to find candidate violations and excludes tasks whose specification only describes task success, user preference, or high-level quality criteria without an enforceable trace-level obligation.
For Skill-Inject, which does not release raw trajectories, we replay its tasks in Claude Code and apply the same trace-level inclusion criteria to the recorded executions.

To construct the benign class, we randomly sample executions from the same benchmark sources and manually inspect them against the corresponding skill specifications.
A randomly sampled execution is retained as benign only when the relevant behavioral obligation is satisfied, is not triggered, or no trace-checkable obligation applies.
We discard sampled executions when the trace lacks the evidence needed to judge the obligation, when the specification is ambiguous, or when the apparent failure is only a task-quality error rather than a behavioral-policy violation.

All retained cases are manually labeled from the skill documentation, the execution trace, and any produced artifacts visible to the monitor.
A trajectory is labeled violating only when the recorded execution contains a concrete event sequence that contradicts the skill's stated behavioral specification.
We use a conservative labeling rule: a case is retained as violating only when the violated obligation and responsible event sequence are both identifiable from the trace; otherwise, the case is excluded rather than adjudicated into either class.
This process produces 152 labeled executions in \textsc{SB+SI}: 72 policy-violating trajectories and 80 benign trajectories.

\section{Baseline Adaptation Protocol}
\label{app:baseline-adaptation}

We keep the policy source and execution trace fixed across all systems, and score every verdict against the same ground-truth label.
For each \textsc{SB+SI} case, AgentSpec and Progent receive the closest faithful rule encoding supported by their action-boundary interfaces.
Policies over the pending action, its arguments, or local trigger conditions are encoded directly.
Policies that require long-range artifact identity, historical data flow, or evidence outside the pending action are encoded as the strongest action-boundary approximation, without adding \tool-style trace memory to the baseline.

The LLM-as-judge baseline receives only the inputs available before enforcement: the full execution trace, the skill specification, and the same policy-relevant context available to \tool.
It returns a binary violation verdict with a short rationale, but does not receive \tool's alert, compiled policy, SMT witness, or the ground-truth label.

\section{Specification-Compilation Prompt Template}
\label{app:prompt-template}

\tool uses the language model only to draft candidate policy documents, not to
judge whether a trace violates a policy. \autoref{fig:prompt-template} shows an abridged template of the prompt interface.

\begin{figure}[H]
\begin{tcblisting}{
  enhanced,
  listing only,
  colback=gray!3,
  colframe=tgModuleLine,
  coltitle=black,
  title={Abridged specification-compilation prompt template},
  fonttitle=\bfseries\footnotesize,
  colbacktitle=tgCompile!45,
  boxrule=0.6pt,
  arc=2pt,
  left=3pt,
  right=3pt,
  top=3pt,
  bottom=3pt,
  listing options={
    basicstyle=\rmfamily\scriptsize\color{black},
    breaklines=true,
    columns=fullflexible
  }
}
Task:
  Translate the natural-language skill policy into a PolicyDocument.
  Return YAML only.

Inputs:
  1. SECURITY_PROTOCOL: the skill's natural-language policy text.
  2. observed_vocabulary: actions, argument roles, command options,
     predicates, scripts, and text samples recovered from the trace.
  3. signature: the allowed condition operators and temporal templates.
  4. policy_language: the target schema and declared role set.

Grounding contract:
  - Draft policy statements only; do not classify the trace.
  - Use only literals and symbols present in the inputs.
  - Do not emit event ids, witness ids, SMT variables, expected verdicts,
    violation labels, or other trace-checking artifacts.
  - Emit Unsupported when a rule cannot be grounded.

Output:
  A YAML PolicyDocument containing policy facts and one or more
  statements over actions, conditions, resources, and temporal templates.
\end{tcblisting}
\caption{Abridged prompt template used for specification compilation.}
\label{fig:prompt-template}
\end{figure}

Every drafted \icode{PolicyDocument} is parsed and validated against the
observed signature before SMT encoding; invalid, ill-typed, or unsupported
clauses are rejected by deterministic components.

\section{Policy Examples}
\label{app:policy-forms}

\autoref{tab:rq1-policy-forms} gives representative policy categories from
\textsc{SB+SI} and their finite-trace formalizations.

\section{The Preset Forms as Instances of \texorpdfstring{$\Box\,\Theta$}{the Fallback}}
\label{app:form-readings}

\autoref{tab:form-readings} writes each preset form of the policy language in the temporal
vocabulary of the fallback.

\begin{table}[H]
\centering
\caption{Each preset form as an instance of the fallback shape $\Box\,\Theta$. Here
$\Diamond^{+}_{\le \ell}$ bounds the search to the next $\ell$ events; ${\sf until}$'s interval condition
is given by its compilation rule.}
\label{tab:form-readings}
\small
\renewcommand{\arraystretch}{1.15}
\begin{tabular}{ll}
\toprule
\textbf{Form} & \textbf{As $\Box\,\Theta$} \\
\midrule
\rowcolor{black!5}
${\sf abs}(\psi)$ & $\Box\,\neg\psi$ \\
${\sf prec}(\psi, \psi')$ & $\Box(\psi \Rightarrow \Diamond^{-}\psi')$ \\
\rowcolor{black!5}
${\sf resp}(\psi, \psi')$ & $\Box(\psi \Rightarrow \Diamond^{+}\psi')$ \\
${\sf bresp}_\ell(\psi, \psi')$ & $\Box(\psi \Rightarrow \Diamond^{+}_{\le \ell}\psi')$ \\
\rowcolor{black!5}
${\sf rslv}(\psi, \psi')$ & $\Box(\psi \Rightarrow \psi' \vee \Diamond^{+}(\psi' \vee \psi))$ \\
${\sf until}(\psi, \psi', \psi_b)$ & after $\psi$: no $\psi_b$ before a resolving $\psi'$ \\
\bottomrule
\end{tabular}
\end{table}

\end{document}